\documentclass[draftclsnofoot,onecolumn, 12pt]{IEEEtran}
\usepackage{graphicx}
\usepackage{epstopdf}
\usepackage{graphicx}
\usepackage{amsmath}
\usepackage{amsfonts}
\usepackage{amssymb, color}
\allowdisplaybreaks
\newtheorem{theorem}{Theorem}

\newtheorem{lemma}{Lemma}

\begin{document}

%-------------------------- paper title--------------------------------------------%
\title{On the Secrecy  Capacity of a Full-Duplex Wirelessly Powered Communication System}
\author{Ivana Nikoloska, \emph{Student Member, IEEE}, Nikola Zlatanov, \emph{Member, IEEE}, Zoran Hadzi-Velkov, \emph{Senior Member, IEEE}, and Rui Zhang, \emph{Fellow, IEEE} \thanks{Ivana Nikoloska and Nikola Zlatanov are with the Department of Electrical and Computer Systems Engineering, Monash University, Melbourne, Australia. Emails: ivana.nikoloska@monash.edu, nikola.zlatanov@monash.edu}  \thanks{Zoran Hadzi-Velkov is with the Faculty of Electrical Engineering and Information Technologies, Ss. Cyril and Methodius University, Skopje, Macedonia. Email: zoranhv@feit.ukim.edu.mk} \thanks{Rui Zhang is with the Department of Electrical and Computer Engineering, National University of Singapore. Email: elezhang@nus.edu.sg} \thanks{This paper, in part, has been accepted for presentation at the 2019 IEEE ISWCS.}}
\maketitle

%-------------------------- abstract--------------------------------------------%
\vspace{-3mm}
\begin{abstract}
In this paper, we investigate the secrecy capacity of a point-to-point, full-duplex (FD) wirelesly powered communication system in the presence of a passive eavesdropper (EVE). The considered system is comprised of an energy transmitter (ET), an energy harvesting user (EHU), and a passive EVE. The ET transmits radio-frequency energy, which is used for powering the EHU as well as for generating interference at the EVE. The EHU uses the energy harvested from the ET to transmit confidential messages back to the ET. As a consequence of the FD mode of operation, both the EHU and the ET are subjected to self-interference, which has different effects at the two nodes. In particular, the self-interference impairs the decoding of the received message at the ET, whilst it serves as an additional energy source at the EHU. For this system model, we derive an upper and a lower bound on the secrecy capacity. For the lower bound, we propose a simple achievability scheme. Our numerical results show significant improvements in terms of achievable secrecy rate when the proposed communication scheme is employed against its half-duplex counterpart, even for practical self-interference values at the ET.
\end{abstract}

%------------------------ Introduction--------------------------------------------%
\section{Introduction}\label{Sec-Intro}
The security of wireless communication is of critical societal interest. Traditionally, encryption has been the primary method which ensures that only the legitimate receiver receives the intended message. Encryption algorithms commonly require that some information, colloquially referred to as a key, is shared only among the legitimate entities in the network. However, key management makes the encryption impractical in architectures such as radio-frequency identification (RFID) networks and sensor networks, since certificate authorities or key distributers are often not available and limitations in terms of computational complexity make the use of standard data encryption difficult \cite{Sec_RFID}, \cite{Sec_sensorNets}. This problem with network security will be increasingly emphasised in the foreseeable future because of paradigms such as the Internet of Things (IoT). The IoT, as a ``network of networks'', will  provide ubiquitous connectivity and information-gathering capabilities to a massive number of communication devices. However, the low-complexity hardware and the severe energy constraints of the IoT devices present unique security challenges. To ensure confidentiality in such networks, exploitation of the physical properties of the wireless channel has become an attractive option \cite{Sec_sensorNets}. Essentially, the presence of fading, interference, and path diversity in the wireless channel can be leveraged in order to degrade the ability of potential intruders to gain information about the confidential messages sent through the wireless channel \cite{Sec_sensorNets}. This approach is commonly known as physical layer security, or alternatively as information-theoretic security \cite{4626059}.

Shannon and Wyner have laid a solid foundation for studying secrecy of many different system models in \cite{SecCap_Shannon},\cite{SecCap_Wyner}, including communication systems powered by energy harvesting (EH), which have attracted significant attention recently \cite{SecCap_WPCNs}, \cite{SecCap_EH}. EH relies on harvesting energy from ambient renewable and environmentally friendly sources such as, solar, thermal, vibration or wind, or, from dedicated energy transmitters. The latter gives rise to wirelesly powered communication networks (WPCNs) \cite{bi2016wireless}. EH is often considered as a suitable supplement to IoT networks, since most IoT nodes have low power requirements on the order of microwatts to milliwatts, which can be easily met by EH. In addition, when paired with physical layer security, WPCNs can potentially offer a secure and ubiquitous operation \cite{liu2018exploiting}. An EH network with multiple power-constrained information sources has been studied in \cite{SecCap_PowerConst}, where the authors derived an exact expression for the probability of a positive secrecy capacity. In \cite{SecCap_MIMO_Ulukus} and \cite{SecCap_MIMO_Ulukus_prim}, the secrecy capacity of the EH Gaussian multiple-input-multiple-output (MIMO) wire-tap channel under transmitter- and receiver-side power constraints has been derived. The secrecy outage probability of a single-input-multiple-output (SIMO) and multiple-input-single-output (MISO) simultaneous wireless information and power transfer (SWIPT) systems were characterized in \cite{SecCap_SIMO_SWIPT} and \cite{liu2013secrecy}-\cite{SecCap_MISO_SWIPT}, respectively. Secrecy in SWIPT systems has also been studied in \cite{liu2017power}, \cite{zhang2016artificial}. Relaying networks with EH in the presence of a passive eavesdropper have been studied in \cite{SecCap_Relay}. Defence methods with EH friendly jammers, have been proposed in \cite{SecCap_Jam} and \cite{SecCap_FD_Jam}, where the secrecy capacity and the secrecy outage probability have been derived.

In addition to physical layer security, another appealing option for networks with scarce resources such as WPCNs, is the full-duplex (FD) mode of operation. Recent results in the literature, e.g., \cite{SecCap_FD_Passive}, \cite{SecCap_FD_Comb}, \cite{zhang2017secrecy}, \cite{tang2017physical}, have shown that it is possible for transceivers to operate in the FD mode by transmitting and receiving  signals simultaneously and in the same frequency band. The FD mode of operation can lead to doubling (or even tripling, see \cite{SecCap_FD_TripleSE}) of the  spectral efficiency of the network in question.

Motivated by these advances in FD communication and the applicability of physical layer security to WPCNs, in this paper, we investigate the secrecy capacity of a FD wirelessly powered communication system.Unlike our prior work in \cite{nikoloska2018capacity} which does not have an eavesdropper and therefore does not consider secrecy constraints, the network in this paper is comprised of an energy transmitter (ET) and an energy harvesting user (EHU) in the presence of a passive eavesdropper (EVE). The  objective in this paper is to investigate the secrecy capacity and develop novel communication schemes capable of achieving non-zero secrecy rates. As a result, the achievability schemes presented in this paper are different than the achievability schemes in \cite{nikoloska2018capacity}. In this system, the ET  sends radio-frequency (RF) energy to the EHU, whereas, the EHU harvests this energy and uses it to transmit confidential information back to the ET. The signal transmitted by the ET serves a second purpose by acting as an interference signal for EVE. Both the ET and the EHU are assumed to operate in the FD mode, hence, both nodes transmit and receive RF signals in the same frequency band and at the same time. As a result, both are subjected to self-interference. The self-interference hinders the decoding of the information signal received from the EHU at the ET. At the EHU, the self-interference resulting from the simultaneous transmission and reception increases the amount of energy that can be harvested by the EHU \cite{FD_SI_recycling}. Meanwhile, EVE is passive and only aims to intercept the confidential message transmitted by the EHU to the ET.
For the considered system model, we derive an upper and a lower bound on the secrecy capacity. Furthermore, we provide a simple achievability scheme for the lower bound on the secrecy capacity. The proposed scheme in this paper is relatively simple and therefore easily applicable in practice in wirelessly powered IoT networks which require secure information transmissions. For example, sensors which are embedded in the infrastructure, like buildings, bridges or the power grid, monitor their environment and generate measurements. The generated measurements often contain sensitive information. An Unmanned Aerial Vehicle (UAV) can fly close to the sensors in order to power them, and then receive the generated data packets from the sensors. The proposed scheme in this paper can be used in this scenario and it will guarantees that such sensitive information will never be intercepted by an illegitimate, third party.

The rest of the paper is organized as follows. Section~II provides the system and channel models. Sections~III and ~IV present the upper and the lower bounds on the secrecy capacity, respectively. In Section~V, we provide numerical results and we conclude the paper in Section~VI. Proofs of theorems/lemmas are provided in the Appendices.

%-------------System Model And Problem Formulation------------------------------------%
\vspace{-3mm}
\section{System Model and Problem Formulation}\label{Sec-Sys}

We consider a system model comprised of an EHU, an ET, and an EVE. In order to improve the spectral efficiency of the considered  system, both the EHU and the ET are assumed to operate in the FD mode, i.e., both nodes transmit and receive RF signals simultaneously and in the same frequency band. Thereby, the EHU receives energy signals from the ET and simultaneously transmits information signals to the ET. Similarly, the ET transmits energy signals to the EHU and simultaneously receives information signals from the EHU. The signal transmitted from the ET also serves as interference to the EVE, and thereby increases its noise floor. Due to the FD mode of operation, both the EHU and the ET are subjected to self-interference, which has opposite effects at the two nodes, respectively. More precisely, the self-interference signal has a negative effect at the ET since it hinders the decoding of the information signal received from the EHU. As a result, the ET should be designed with a self-interference suppression apparatus, which can suppress the self-interference at the ET and thereby  improve the decoding of the desired signal received from the EHU.  On the other hand, at the EHU, the self-interference signal is desired since it increases the amount of energy that can be harvested by the EHU.  Hence, the EHU should be designed without  a self-interference suppression apparatus  in order for the energy contained in the self-interference signal to be harvested, i.e., the EHU should perform  energy recycling  as proposed in \cite{FD_SI_recycling}. Meanwhile, EVE remains passive and only receives, thus it is not subjected to self-interference.

\vspace{-3mm}
\subsection{Channel Model}

Let $V_{12i}$ and $V_{21i}$ denote random variables (RVs) which model the fading  channel gains of the  EHU-ET and  ET-EHU channels in channel use $i$, respectively. Due to the FD mode of operation, the EHU-ET and the ET-EHU channels are identical and as a result the channel gains $V_{12i}$ and $V_{21i}$ are assumed to be identical, i.e., $V_{12i}$ $=$ $V_{21i}$ $=$ $V_{i}$. Moreover, let $F_i$ and $G_i$ denote RVs which model the fading channel gains of the EHU-EVE and ET-EVE channels in channel use $i$, respectively. We assume that all channel gains follow a block-fading model, i.e., they remain constant during all channel uses in one block, but change from one block to the next, where each block consists of (infinitely) many channel uses. All three nodes are assumed to have a priori knowledge of the CSI of the EHU-ET channel, i.e., $V_i$. In addition, EVE is assumed to have CSI of the the EHU-EVE and the ET-EVE channels, given by $G_i$ and $F_i$, respectively.

In the $i$-th channel use, let the transmit symbols at the EHU and the ET be modeled as RVs, denoted by $X_{1i}$ and $X_{2i}$, respectively. Moreover, in channel use $i$, let the received symbols at the EHU, the ET, and EVE be modeled as RVs, denoted by $Y_{1i}$, $Y_{2i}$, and $Y_{3i}$, respectively. Furthermore, in channel use $i$, let the RVs modeling the AWGNs at the   EHU, the ET, and the EVE be denoted by $N_{1i}$, $N_{2i}$, and $N_{3i}$, respectively, such that $N_{1i}\sim\mathcal{N} \left(0, \sigma_1^2\right)$, $N_{2i}\sim\mathcal{N} \left(0, \sigma_2^2\right)$, and $N_{3i}\sim\mathcal{N} \left(0, \sigma_3^2\right)$, where $\mathcal{N} \left(\mu, \sigma^2\right)$ denotes a Gaussian distribution with mean $\mu$ and variance $\sigma^2$. Note that the derived results can be extended to the complex-valued channel in a straightforward manner since the complex-valued channel can be resolved in two parallel real-valued channels that use the achievability schemes that we derive here. Moreover, let the RVs modeling the additive self-interferences at the EHU and the ET in channel use $i$ be denoted by $I_{1i}$ and $I_{2i}$, respectively.

By using the notation defined above, the input-output relations describing the considered channel in channel use $i$ can be written as
\begin{align}
Y_{1i}&=V_{i} X_{2i}+I_{1i}+N_{1i},\label{eq_1i}\\
Y_{2i}&=V_{i} X_{1i}+I_{2i}+N_{2i},\label{eq_2i}\\
Y_{3i}&=F_{i} X_{1i}+G_{i} X_{2i}+N_{3i}.\label{eq_2ia}
\end{align}

\vspace{-3mm}
\subsection{Self-Interference Model}
A general model for the self-interference at the EHU and the ET is given by \cite{SecCap_FDradios}
\begin{align}
I_{1i}&=\sum_{m=1}^{M} \tilde Q_{1,m}(i) X_{1i}^m,\label{eq_si_1i}\\
I_{2i}&=\sum_{m=1}^{M} \tilde Q_{2,m}(i) X_{2i}^m,\label{eq_si_2i}
\end{align}
where $M < \infty$ is an integer and $\tilde Q_{1,m}(i)$ and $\tilde Q_{2,m}(i)$ model the $m$-th component of the self-interference channel between the transmitter- and the receiver-ends at the EHU and the ET in channel use $i$, respectively. As shown in \cite{SecCap_FDradios}, the components in (\ref{eq_si_1i}) and (\ref{eq_si_2i}) for which $m$ is odd carry non-negligible energy and the remaining components carry negligible energy and therefore can be ignored. Furthermore, the higher order components carry less energy than the lower order terms. As a result, we can justifiably adopt the first-order approximation of the self-interference in (\ref{eq_si_1i}) and (\ref{eq_si_2i}), and model $I_{1i}$ and $I_{2i}$ as
\begin{align}
I_{1i}&=\tilde Q_{1i} X_{1i},\label{eq_si_3i}\\
I_{2i}&=\tilde Q_{2i} X_{2i},\label{eq_si_4i}
\end{align}
where $\tilde Q_{1i}=\tilde Q_{1}(i)$ and $\tilde Q_{2i}=\tilde Q_{2}(i)$ are used for simplicity of notation. Thereby, the adopted self-interference model takes into account only the linear component of (\ref{eq_si_1i}) and (\ref{eq_si_2i}), i.e., the component for $m=1$. The linear self-interference model has been widely used, e.g. in \cite{SecCap_FDradios}, \cite{SecCap_zlatanov}.

By inserting (\ref{eq_si_3i}) and (\ref{eq_si_4i}) into (\ref{eq_1i}) and (\ref{eq_2i}), respectively, we obtain
\begin{align}
Y_{1i}&=V_{i} X_{2i}+\tilde Q_{1i} X_{1i}+N_{1i},\label{eq_1a}\\
Y_{2i}&=V_{i} X_{1i}+\tilde Q_{2i} X_{2i}+N_{2i}. \label{eq_2a}
\end{align}

To model the worst-case of linear self-interference, we note the following. Since the ET knows which symbol it has transmitted in channel use $i$, the ET knows the outcome of the RV $X_{2i}$, denoted by $x_{2i}$. As a result of this knowledge, the noise that the ET ``sees'' in its received symbol $Y_{2i}$ given by (\ref{eq_2a}), is $\tilde Q_{2i} x_{2i}+N_{2i}$, where $x_{2i}$ is a constant. Hence, the noise that the ET ``sees'', $\tilde Q_{2i} x_{2i}+N_{2i}$, will represent the worst-case of noise, under a second moment constraint, if and only if $\tilde Q_{2i}$ is an independent and identically distributed (i.i.d.) Gaussian RV\footnote{This is due to the fact that the Gaussian distribution has the largest entropy under a second moment constraint, see \cite{cover2012elements}.}.  Therefore, in order to derive results for the worst-case of linear self-interference, we assume that $\tilde Q_{2i}\sim\mathcal{N}\{\bar q_{2i},\alpha_2\}$ in the rest of the paper. Meanwhile, $Q_{1i}$ is distributed according to an arbitrary probability distribution with mean $\bar q_{1i}$ and variance $\alpha_1$.

Now, since $\tilde Q_{1i}$ and $\tilde Q_{2i}$ can be written equivalently as $\tilde Q_{1i} = Q_{1i} + \bar q_1$ and $\tilde Q_{2i} = Q_{2i} + \bar q_2$, where $\bar q_{1i}$ and $\bar q_{2i}$ are the means of $\tilde Q_{1i}$ and $\tilde Q_{2i}$, respectively, and $Q_{1i}$ and $Q_{2i}$ denote the remaining zero-mean components of $\tilde Q_{1i}$ and $\tilde Q_{2i}$, respectively, we can write $Y_{1i}$ and $Y_{2i}$ in (8) and (\ref{eq_2a}), respectively, as
\begin{align}
Y_{1i}&=V_{i} X_{2i}+\bar q_{1i} X_{1i} + Q_{1i} X_{1i}+N_{1i},\label{eq_3a}\\
Y_{2i}&=V_{i} X_{1i}+\bar q_{2i} X_{2i} + Q_{2i} X_{2i}+N_{2i}. \label{eq_4a}
\end{align}

Since the ET always knows the outcome of $X_{2i}$, $x_{2i}$, and since given sufficient time it can always estimate the deterministic component of its self-interference channel, $\bar {q_2}$, the ET can remove $\bar {q_2} X_{2i}$ from its received symbol $Y_{2i}$, given by (\ref{eq_4a}), and thereby reduce its self-interference. In this way, the ET obtains a new received symbol, denoted again by $Y_{2i}$, as
\begin{align} \label{eq_5a1}
Y_{2i}&=V_{i} X_{1i}+   Q_{2i} X_{2i}+N_{2i}.
\end{align}
Note that, $X_{2i}$ travels from the transmitter-side to the receiver-side at the ET, it undergoes a random transformation which is unknown to the ET. Thereby, since $Q_{2i}$ in (\ref{eq_5a1}) changes independently from one channel use to the next, the ET cannot estimate and remove  $Q_{2i} X_{2i}$ from its received symbol even though the ET knows the outcome of $X_{2i}$. Thus, $Q_{2i} X_{2i}$ in (\ref{eq_5a1}) is the residual self-interference at the ET where the ET knows the outcome of $X_{2i}$. On the other hand, since the EHU benefits from the self-interference, it does not remove  $\bar q_1 X_{1i}$ from its received symbol  $Y_{1i}$, given by (\ref{eq_3a}), in order to have a self-interference signal with a much higher energy, which it can then harvest. Hence, the received symbol at the EHU is given by (\ref{eq_3a}).

In this paper, we are interested in the secrecy capacity of the channel characterised by the input-output relationships given by (\ref{eq_3a}), (\ref{eq_5a1}), and (\ref{eq_2ia}).

\vspace{-3mm}
\subsection{Energy Harvesting Model}
The energy harvested by the EHU in channel use $i$ is given by \cite{FD_SI_recycling}
\begin{align}\label{eq_3energy}
%E_i=\text{max} \{ 0, \eta (H_{i}X_{2i} + \bar g_{1} X_{1i}+G_{1i} X_{1i})^2-P_p \},
E_{{\rm in,}i}= \eta (V_{i}X_{2i} + \bar q_{1} X_{1i}+Q_{1i} X_{1i})^2,
\end{align}
where $0<\eta<1$ is the energy harvesting efficiency coefficient. For convenience, we have assumed unit time and thus we use the terms power and energy interchangeably in the sequel. The EHU stores $E_{{\rm in,}i}$ in its battery, which is assumed to have an infinitely large storage capacity. Let $B_i$ denote the amount of harvested energy in the battery of the EHU at the end of the $i$-th channel use. Moreover, let $E_{{\rm out,}i}$ be the extracted energy from the battery in the $i$-th channel use. Then, $B_i$, can be written as
\begin{align}\label{eq_4energy}
B_i=B_{i-1}+E_{{\rm in,}i}-E_{{\rm out,}i}.
\end{align}
Since in channel use $i$ the EHU cannot extract more energy than the amount of energy stored in its battery at the end of channel use $i-1$, the extracted energy from the battery in channel use $i$,  $E_{{\rm out,}i}$, can be obtained as
\begin{align}\label{eq_4aa}
E_{{\rm out,}i}=\min\{B_{i-1},X_{1i}^2 + P_p\},
\end{align}
where   $X_{1i}^2$ is the  transmit energy of the desired transmit symbol in channel use $i$, $X_{1i}$,  and $P_p$ is the processing energy cost of the EHU \cite{xu2014throughput}. The processing cost, $P_p$, models the system level power consumption at the EHU, i.e., the energy spent due to the electrical components in the electrical circuit such as AC/DC convertors and RF amplifiers as well as the energy spent for processing. Note that the ET also requires energy for processing. However, the ET is assumed to be equipped with a conventional power source which is always capable of providing the processing energy  without affecting the energy required for transmission.

Now, if the total number of channel uses satisfies $n \rightarrow \infty$, if the battery of the EHU has an unlimited storage capacity, and furthermore
\begin{align}\label{eq_4b}
\mathcal{E}\{E_{{\rm in,}i}\}\geq \mathcal{E}\{X_{1i}^2\}+P_p
\end{align}
 holds, where $\mathcal{E}\{\cdot\}$ denotes statistical expectation,  then the number of channel uses in which the extracted energy from the battery is insufficient  and thereby  $E_{{\rm out,}i}=B_{i-1}$ holds is negligible compared to the number of channel uses in which the extracted energy is sufficient for both transmission and processing \cite{eh_zlatanov}. In other words, when the above three conditions hold, in almost all channel uses, there will be enough energy to be extracted from the EHU's battery for both processing, $P_p$, and for the transmission of the desired transmit symbol $X_{1i}$, $X_{1i}^2$, and thereby $E_{{\rm out,}i}=X_{1i}^2+P_p$ holds.

\vspace{-3mm}
\section{Upper Bound on the Secrecy Capacity}\label{Cap}
For the considered channel, we propose the following theorem which establishes an upper bound on the secrecy capacity.
\begin{theorem}
Assuming that the average power constraint at the ET is $P_{ET}$, an upper bound on the secrecy capacity of the considered channel is given by
%\newpage
%\begin{footnotesize}
\begin{align}\notag
& \underset{p(x_1|x_2,v),p(x_2|v)}{\text{max}} \sum_{x_2 \in \mathcal{X}_2} \sum_{v \in \mathcal{V}}I(X_1;Y_2 | X_2=x_2, V=v)p(x_2|v)p(v)\nonumber\\
&\qquad\qquad\qquad-\sum_{v \in \mathcal{V}}\sum_{g \in \mathcal{G}}\sum_{f \in \mathcal{F}}I(X_1;Y_3 | V=v, G=g, F=f)p(v)p(g)p(f) \nonumber\\
&{\rm{Subject\;\;  to}}  \nonumber\\
&\qquad\qquad{\rm C1:} \sum_{x_2 \in \mathcal{X}_2} \sum_{v \in \mathcal{V}} x_2^2 p(x_2|v)p(v) \leq P_{ET}  \nonumber\\
&\qquad\qquad{\rm C2:} \int_{x_1} \sum_{x_2 \in \mathcal{X}_2} \sum_{v \in \mathcal{V}}(x_1^2 + P_p) p(x_1|x_2,v)p(x_2|v)p(v)dx_1 \leq  \nonumber\\
&\qquad\qquad\qquad \int_{x_1} \sum_{x_2 \in \mathcal{X}_2} \sum_{v \in \mathcal{V}}E_{\rm in} p(x_1|x_2,v)p(x_2|v)p(v)dx_1  \nonumber\\
&\qquad\qquad{\rm C3:} \sum_{x_2 \in \mathcal{X}_2}{p(x_2|v)} = 1  \nonumber\\
&\qquad\qquad{\rm C4:} \int_{x_1}{p(x_1|x_2,v) d x_1} = 1,    \label{cap_eq1}
\end{align}
%\end{footnotesize}
where $I(;|)$ denotes the conditional mutual information. In (\ref{cap_eq1}), lower-case letters $x_2$, $v$, $g$, and $f$ represent realizations of the random variables $X_2$, $V$, $G$, and $F$, respectively, and their support sets are denoted by $\mathcal{X}_2$, $\mathcal{V}$, $\mathcal{G}$, and $\mathcal{F}$, respectively. Constraint C1 in  (\ref{cap_eq1}) constrains the average transmit power of the ET to $P_{ET}$, and C2 is due to (\ref{eq_4b}), i.e., due to the fact that EHU has to have harvested enough energy for both processing and transmission of symbol $X_1$. The maximum in the objective function is taken over all possible conditional probability distributions of $x_1$ and $x_2$, given by $p(x_1|x_2,v)$ and $p(x_2|v)$, respectively.
\end{theorem}
\begin{IEEEproof}
Please refer to Appendix~\ref{Con}, where the converse is provided.
\end{IEEEproof}

\vspace{-3mm}
\subsection{Simplified Expression of the Upper Bound on the Secrecy Capacity}
The optimal input distributions at the EHU and the ET that are the solutions of the optimization problem in (\ref{cap_eq1}) and the resulting simplified expressions of the upper bound on the secrecy capacity are provided by the following lemma.
\begin{lemma}
The optimal input distribution at the EHU, found as the solution of the optimization problem in (\ref{cap_eq1}), is zero-mean Gaussian with variance $P_{EHU}(x_2,v)$, i.e., $p(x_1|x_2,v) \sim\mathcal{N} \left(0, P_{EHU}(x_2,v)\right)$, where $P_{EHU}(x_2,v)$ can be found as the solution of
%\begin{footnotesize}
\begin{align}\label{cap_eq1aa}
&\frac{v^2}{\sigma_2^2 + x_{2}^2 \alpha_2} +\left(1+\frac{v^2 P_{EHU}(x_2,v)}{\sigma_2^2 + x_2^2 \alpha_2}\right) \sum_{f \in \mathcal{F}} \frac{f^2}{f^2 P_{EHU}(x_2,v)+\sigma_3^2} p(f)\nonumber\\
=&\left(1+\frac{v^2 P_{EHU}(x_2,v)}{\sigma_2^2 + x_2^2 \alpha_2}\right)\lambda_2 (1-\eta(\bar {g_1}^2 +\alpha_1)),
\end{align}
%\end{footnotesize}
where $\lambda_2$ is chosen such that
%\begin{align} \label{cap_lambda}
% &\left(1-\eta(\bar g_1^2 +\alpha_1)\right)\sum_{v \in \mathcal{V}}\sum_{x_2 \in \mathcal{X}_2} P_{EHU}(X_2,V) \left(x_{2j},h\right)p(x_2|v) p(v)+P_p =\nonumber\\
% &\eta P_{ET} \sum_{v \in \mathcal{V}} \sum_{x_2 \in \mathcal{X}_2} h^2 x_{2j}^2 p(x_2|v) p(v)
%\end{align}
C2 in (\ref{cap_eq1}) holds with equality.

On the other hand, the optimal input distribution at the ET, found as the solution of the optimization problem in (\ref{cap_eq1}), has the following discrete form
\begin{align}\label{cap_eq22a}
p(x_2|v)=p (x_2 = 0) \delta(x_2) + \frac{1}{2} \sum_{j=1}^J  p (x_2 = x_{2j}) \Big(\delta(x_2-x_{2j})+\delta(x_2+x_{2j})\Big),
\end{align}
where $\delta(\cdot)$ denotes the Dirac delta function. Finally, the simplified expression of the upper bound on the secrecy capacity in (\ref{cap_eq1}), denoted by $C_s^u$, is given by
\begin{align}
&C_s^u=\frac {1}{2}\sum_{v \in \mathcal{V}} \sum_{j=1}^{J} \log \left(1+\frac{v^2 P_{EHU}(x_2,v)}{\sigma_2^2 + x_{2j}^2 \alpha_2}\right)p (x_2 = x_{2j}) p(v)\nonumber\\
&+ \sum_{v \in \mathcal{V}}\sum_{g \in \mathcal{G}}\sum_{f \in \mathcal{F}} \left[ \int_{-\infty}^{\infty} \frac{1}{\sqrt{2 \pi \sigma_{y_3}^2}} \sum_{j=1}^{J} p (x_2 = x_{2j}) e^{-\frac{(y_3 - x_{2j})^2}{2 \sigma_{y_3}^2}} \right.\nonumber\\
&\left.\times \ln \left(\frac{1}{\sqrt{2 \pi \sigma_{y_3}^2}} \sum_{j=1}^{J} p (x_2 = x_{2j}) e^{-\frac{(y_3 - x_{2j})^2}{2 \sigma_{y_3}^2}}\right) d y_3  \right. -\int_{-\infty}^{\infty} \frac{1}{\sqrt{2 \pi \sigma_{3}^2}} \left. \sum_{j=1}^{J} p (x_2 = x_{2j}) e^{-\frac{(z - x_{2j})^2}{2 \sigma_3^2}} \right.\nonumber\\
&\left.\times \ln \left(\frac{1}{\sqrt{2 \pi \sigma_{3}^2}} \sum_{j=1}^{J} p (x_2 = x_{2j}) e^{-\frac{(z - x_{2j})^2}{2 \sigma_3^2}}\right) d z_3 \right] p(v)p(g)p(f) \LARGE{\label{cap_eq3}}.
\end{align}
%\end{footnotesize}
\end{lemma}
%\begin{equation}\label{cap_eq3}
%  1
%\end{equation}
\begin{IEEEproof}
Please refer to Appendix B.
\end{IEEEproof}

\vspace{-3mm}
\section{Lower Bound on the Secrecy Capacity - An Achievable Secrecy Rate}\label{LowCap}
From Lemma 1, we can see that the upper bound on the secrecy capacity cannot be achieved since the EHU has to know $x_{2i}^2$ in each channel use $i$, in order for the EHU to calculate (\ref{cap_eq1aa}). In other words, the EHU can not adapt $P_{EHU}(x_2,v)$ and the data rates of its codewords accordingly. The knowledge of $x_{2i}^2$ at the EHU is not possible since the input distribution at the ET, given by (\ref{cap_eq22a}), is discrete with a finite number of probability mass points. However, if we set the input distribution at the ET to be binary such that $x_{2i}$, $\forall i$, takes values from the set $\{x_0,-x_0\}$, then the EHU can know $x_{2i}^2$ in each channel use $i$ since $x_{2i}^2 = x_0^2$ , $\forall i$, and therefore this rate can be achieved. Hence, to obtain an achievable lower bound on the secrecy capacity, we propose the ET to use the following input distribution
\begin{align}\label{cap_eq22a_low}
p(x_2|v)=\frac{1}{2} \Big(\delta(x_2-x_{0}(v))+\delta(x_2+x_{0}(v))\Big).
\end{align}
The value of $x_0(v)$ will be determined in the following.

\vspace{-3mm}
\subsection{Simplified Expression of the Lower Bound on the Secrecy Capacity}
The simplified expression for the lower bound on the secrecy capacity resulting from the ET using the distribution given by (\ref{cap_eq22a_low}), is provided by the following lemma.

\begin{lemma}
Let us define $\mathcal{I} (x)$ as
\begin{align} \label{lem}
\mathcal{I} (x) = \frac{2}{\sqrt{2 \pi} x} e^{-x^2/2} \int_{0}^{\infty} e^{-y^2 / 2x} \cosh(y) \ln (\cosh(y)) dy.
\end{align}

Depending on the channel qualities, we have three cases for the achievable secrecy rate.

\textit{Case 1:} If the following conditions hold
%\begin{footnotesize}
\begin{align}\label{cap_low_cond_1}
&\frac{1}{2} \sum_{v \in \mathcal{V}}\log \left(1+\frac{v^2 P_{EHU}(x_2,v)}{\sigma_2^2 + P_{ET} \alpha_2}\right) p(v) +\lambda_1 P_{ET} \nonumber\\
=& \lambda_2 \left((1-\eta(\bar {q_1}^2 +\alpha_1))\sum_{v \in \mathcal{V}}P_{EHU}(x_2,v)p(v)- \eta P_{ET} \Omega_V \right),
\end{align}
%\end{footnotesize}
and
%\newpage
%\begin{footnotesize}
\begin{align} \label{cond_1b}
&\frac{1}{2} \sum_{v \in \mathcal{V}}\log \left(1+\frac{v^2 P_{EHU}(x_2,v)}{\sigma_2^2 + P_{ET} \alpha_2}\right) p(v) > \sum_{v \in \mathcal{V}}\sum_{g \in \mathcal{G}}\sum_{f \in \mathcal{F}} \left[\frac{1}{2} \ln \left(2 \pi e \sigma_{y_3}^2\right) \right. \nonumber\\
&\left. + \frac{P_{ET}}{f^2 P_{EHU} (x_2, v) + \sigma_3^2} - \mathcal{I} \left(\frac{\sqrt{P_{ET}}}{\sqrt{f^2 P_{EHU} (x_2, v) + \sigma_3^2}}\right) \right. \nonumber\\
&\left. - \frac{1}{2} \ln \left(2 \pi e \sigma_{3}^2\right) - \frac{P_{ET}}{\sigma_3^2} + \mathcal{I} \left(\frac{\sqrt{P_{ET}}}{\sigma_3}\right) \right] p(v)p(g)p(f),
\end{align}
%\end{footnotesize}
where $P_{EHU}(x_2,v)$ is the root of (\ref{cap_eq1aa}) for $x_2=\sqrt{P_{ET}}$ and $\Omega_V=\mathcal{E} \{ v^2 \}$, then the input distribution at the ET has the following form
%\begin{footnotesize}
\begin{align}\label{cap_eq22a_low_1}
p(x_2|v)=\frac{1}{2} \left(\delta\left(x_2-\sqrt{P_{ET}}\right)+\delta \left(x_2+\sqrt{P_{ET}}\right)\right), \forall v.
\end{align}
%\end{footnotesize}
On the other hand, the input distribution at the EHU is zero-mean Gaussian with variance $P_{EHU}(\sqrt{P_{ET}},v)$, i.e., $p(x_1|x_2,v) \sim\mathcal{N} \left(0, P_{EHU}(\sqrt{P_{ET}},v)\right)$, where $P_{EHU}(\sqrt{P_{ET}},v)$ can be found as the solution of (\ref{cap_eq1aa}) for $x_2=\sqrt{P_{ET}}$.

For Case 1, the achievable secrecy rate, denoted by $C_s^l$, is given by
\begin{align} \label{cap_eq3_low}
\hspace{-1.5mm}&C_s^l=\frac {1}{2}\sum_{v \in \mathcal{V}} \log \left(1+\frac{v^2 P_{EHU}(\sqrt{P_{ET}},v)}{\sigma_2^2 + P_{ET} \alpha_2}\right)p(v)\nonumber\\
&\hspace{-1.5mm}+ \sum_{v \in \mathcal{V}}\sum_{g \in \mathcal{G}}\sum_{f \in \mathcal{F}} \left[ \int_{-\infty}^{\infty} \frac{1}{2 \sqrt{2 \pi \sigma_{y_3}^2}}\left( e^{-\frac{(y_3 - \sqrt{P_{ET}})^2}{2 \sigma_{y_3}^2}} + e^{-\frac{(y_3 + \sqrt{P_{ET}})^2}{2 \sigma_{y_3}^2}} \right)\right.\nonumber\\
&\hspace{-1.5mm}\left.\times \ln \left(\frac{1}{2\sqrt{2 \pi \sigma_{y_3}^2}} \left( e^{-\frac{(y_3 - \sqrt{P_{ET}})^2}{2 \sigma_{y_3}^2}} + e^{-\frac{(y_3 + \sqrt{P_{ET}})^2}{2 \sigma_{y_3}^2}} \right)\right) d y_3  \right.\nonumber\\
&-\int_{-\infty}^{\infty}\left. \frac{1}{2\sqrt{2 \pi \sigma_{3}^2}}\left( e^{-\frac{(z_3 - \sqrt{P_{ET}})^2}{2 \sigma_{3}^2}} + e^{-\frac{(z_3 + \sqrt{P_{ET}})^2}{2 \sigma_{3}^2}} \right)\right.\nonumber\\
&\hspace{-1.5mm} \left. \times \ln \left( e^{-\frac{(z_3 - \sqrt{P_{ET}})^2}{2 \sigma_{3}^2}} + e^{-\frac{(z_3 + \sqrt{P_{ET}})^2}{2 \sigma_{3}^2}} \right) d z_3 \right] p(v)p(g)p(f).
\end{align}
%\end{footnotesize}

\textit{Case 2:} If (\ref{cap_low_cond_1}) does not hold, and
%\begin{footnotesize}
\begin{align} \label{cond_1c}
&\frac{1}{2} \sum_{v \in \mathcal{V}}\log \left(1+\frac{v^2 P_{EHU}(x_2,v)}{\sigma_2^2 + x_0^2(v) \alpha_2}\right) p(v) > \sum_{v \in \mathcal{V}}\sum_{g \in \mathcal{G}}\sum_{f \in \mathcal{F}} \left[\frac{1}{2} \ln \left(2 \pi e \sigma_{y_3}^2\right) \right. \nonumber\\
&+ \left.\frac{x_0^2(v)}{f^2 P_{EHU} (x_2, v) + \sigma_3^2} - \mathcal{I} \left(\frac{x_0(v)}{\sqrt{f^2 P_{EHU} (x_2, v) + \sigma_3^2}}\right) \right. \nonumber\\
&\left. - \frac{1}{2} \ln \left(2 \pi e \sigma_{3}^2\right) - \frac{x_0^2(v)}{\sigma_3^2} + \mathcal{I} \left(\frac{x_0(v)}{\sigma_3}\right) \right] p(v)p(g)p(f)
\end{align}
%\end{footnotesize}
holds, then the input distribution at the ET is given by
%\begin{footnotesize}
\begin{align}\label{cap_low_cond_2}
p(x_2|v)=\frac{1}{2} \Big(\delta(x_2-x_0(v))+\delta(x_2+x_0(v))\Big),
\end{align}
%\end{footnotesize}
%
whereas the input distribution at the EHU is zero-mean Gaussian with variance $P_{EHU}(x_0(v),v)$. In this case, $P_{EHU}(x_0(v),v)$ and $x_0(v)$ are the roots of the system of equations comprised of (\ref{cap_eq1aa}) for $x_2=x_0(v)$ and the following equation
%\begin{footnotesize}
\begin{align}\label{cap_low_cond_3}
&\frac{1}{2} \log \left(1+\frac{v^2 P_{EHU}(x_0(v),v)}{\sigma_2^2 + x_0^2(v) \alpha_2}\right) - \lambda_1 x_0^2(v) \nonumber\\
=&\lambda_2 \left((1-\eta(\bar {q_1}^2 +\alpha_1))P_{EHU}(x_0(v),v)- \eta v^2 x_0^2(v) \right).
\end{align}
%\end{footnotesize}

For Case 2, the achievable secrecy rate is given by
%\begin{footnotesize}
\begin{align} \label{cap_low_cond_4}
&C_s^l=\frac {1}{2}\sum_{v \in \mathcal{V}} \log \left(1+\frac{v^2 P_{EHU}(x_0(v),v)}{\sigma_2^2 + x_0^2(v) \alpha_2}\right)p(v)\nonumber\\
&+ \sum_{v \in \mathcal{V}}\sum_{g \in \mathcal{G}}\sum_{f \in \mathcal{F}} \left[ \int_{-\infty}^{\infty} \frac{1}{2 \sqrt{2 \pi \sigma_{y_3}^2}}\left( e^{-\frac{(y_3 - x_0(v))^2}{2 \sigma_{y_3}^2}} + e^{-\frac{(y_3 + x_0(v))^2}{2 \sigma_{y_3}^2}} \right)\right.\nonumber\\
&\left.\times \ln \left(\frac{1}{2\sqrt{2 \pi \sigma_{y_3}^2}} \left( e^{-\frac{(y_3 - x_0(v))^2}{2 \sigma_{y_3}^2}} + e^{-\frac{(y_3 + x_0(v))^2}{2 \sigma_{y_3}^2}} \right)\right) d y_3  \right.\nonumber\\
&-\int_{-\infty}^{\infty}\left. \frac{1}{2\sqrt{2 \pi \sigma_{3}^2}}\left( e^{-\frac{(z_3 - x_0(v))^2}{2 \sigma_{3}^2}} + e^{-\frac{(z_3 + x_0(v))^2}{2 \sigma_{3}^2}} \right)\right.\nonumber\\
& \left. \times \ln \left( e^{-\frac{(z_3 - x_0(v))^2}{2 \sigma_{3}^2}} + e^{-\frac{(z_3 + x_0(v))^2}{2 \sigma_{3}^2}} \right) d z_3 \right] p(v)p(g)p(f).
\end{align}
%\end{footnotesize}

\textit{Case 3:} If neither (\ref{cond_1b}) nor (\ref{cond_1c}) holds, then, the achievable secrecy rate is $C_s^l=0$.
\end{lemma}
%\begin{equation}\label{cap_eq3}
%  1
%\end{equation}
\begin{IEEEproof}
In order for C1 in (\ref{cap_eq1}) to hold, or equivalently for C1 in (\ref{appB_eq0a}) to hold, there are two possible cases for $x_2$. In Case 1, C1 in (\ref{appB_eq0a}) is satisfied if $x_{2}$ is set to take values from the set $\{\sqrt{P_{ET}},-\sqrt{P_{ET}}\}$. If (\ref{appB_eq2}) for $x_2^2=P_{ET}$ does not hold, then $x_2$ is set to take values from the set $\{x_0(v), -x_0(v)\}$, where $x_0(v)$ is given by (\ref{cap_low_cond_3}) in order for C1 in (\ref{appB_eq0a}) to be satisfied. Now, since $C_s^l=\sum_{x_2 \in \mathcal{X}_2} \sum_{v \in \mathcal{V}}I(X_1;Y_2 | X_2=x_2, V=v)p(x_2|v)p(v)-\sum_{v \in \mathcal{V}}\sum_{g \in \mathcal{G}}\sum_{f \in \mathcal{F}}I(X_1;Y_3 | V=v, G=g, F=f)p(v)p(g)p(f)$, where $X_1$ follows a Gaussian probability distribution, and $X_2$ is distributed according to (\ref{cap_eq22a_low_1}) and (\ref{cap_low_cond_2}) for Case 1 and Case 2, respectively, we obtain the expressions in (\ref{cap_eq3_low}) and (\ref{cap_low_cond_4}) by using (\ref{appB_eq0a}) and (\ref{proof_eq4}).
\end{IEEEproof}

Lemma 2 gives insights into the achievability scheme of the derived lower bound on the secrecy capacity. When Case 1 of Lemma 2 holds, the achievability scheme is very simple. In particular, the ET only chooses between $-\sqrt{P_{ET}}$ or $\sqrt{P_{ET}}$ in every channel use. When Case 2 of Lemma 2 holds, from (\ref{cap_low_cond_3}) we see that the ET adapts its transmit power to the channel fading states of the EHU-ET channel, $v$, and increases its transmit power when $v$ is larger, and conversely, it lowers its transmit power when $v$ is not as favourable. As for the EHU, we first note that, since the EHU knows the square of the transmit symbol of the ET $x_2$ in a given channel use, the EHU can adapt its transmit power and its rate in the given channel use according to the expected self-interference at the ET, which depends on the value of $x_2^2$. Secondly, the EHU also takes advantage of the better channel fading states of the EHU-ET channel, $v$, and increases its transmit power and rate when $v$ is larger, and conversely, it lowers its transmit power and rate when $v$ is not as strong. Thirdly, since $\lambda_2$ is chosen such that constraint C2 in (\ref{cap_eq1}) holds, the transmit power of the EHU $P_{EHU}(x_2,v)$ depends on the processing cost $P_p$. Thereby, when Case 2 holds, the ET also takes into account the processing cost of the EHU. 
\vspace{-3mm}
\subsection{Achievability of the Lower Bound on the Secrecy Capacity}\label{Ach}
We set the total number of channel uses (i.e., symbols) $n$ to $n= k (N+B)$, where $N+B$ denotes the total number of time slots used for the transmission and $k$ denotes the number of symbols transmitted per time slot, where $n \to \infty$, $k \to \infty$, $N \to \infty$, and $(N+B) \to \infty$.

Let $\mathcal{N}$ denote a set comprised of the time slots during which the EHU has enough energy harvested and thereby transmits a codeword, and let $\mathcal{B}$ denote a set comprised of the time slots during which the EHU does not have enough energy harvested and thereby it is silent. Let $N=|\mathcal{N}|$ and  $B=|\mathcal{B}|$, where $|\cdot|$ denotes the cardinality of a set.

\emph{Transmissions at the ET:} During the $k$ channel uses of a considered time slot with fading realisation $v$, the ET's transmit symbol is drawn from the probability distribution given in Lemma 2. Thus, in each channel use of the considered time slot, the ET transmits either $\sqrt{P_{ET}}$ or $-\sqrt{P_{ET}}$ with probability $1/2$ if Case 1 in Lemma 2 holds, or transmits $x_0(v)$ or $-x_0(v)$ with probability $1/2$ if Case 2 in Lemma 2 holds.

\emph{Reception of Energy and Transmission of Information at the EHU:} The EHU first generates all binary sequences of length $kNR_{EHU}$, where
\begin{align}\label{eqn0}
R_{EHU}=\frac {1}{2}\sum_{v \in \mathcal{V}} \log \left(1+\frac{v^2 P_{EHU}(x_2,v)}{\sigma_2^2 + x_2^2 \alpha_2}\right)p(v),
\end{align}
where $P_{EHU}(x_2,v)$ and $x_2$ can be found from Lemma 2 depending on which case holds. Then the EHU uniformly assigns each generated sequence to one of $2^{k N R_s}$ groups, where $R_s$ is given by (\ref{cap_eq3_low}) for  Case 1 of Lemma 2, or by (\ref{cap_low_cond_4}) for Case 2 of Lemma 2. The confidential message $W$ drawn uniformly from the set $W \in \{1,\,2,\,...,2^{kNR_s}\}$ is then assigned to a group. Next, the EHU randomly select a binary sequence from the corresponding group to which $W$ is assigned, according to the uniform distribution. This binary sequence is then mapped to a codeword comprised of $kN$ symbols, which is to be transmitted in $N+B$ time slots.  The symbols of the codeword are drawn according to a zero-mean, unit-variance Gaussian distribution. Next, the codeword is divided into $N$ blocks, where each block is comprised of $k$ symbols. The length of each block is assumed to coincide with a single fading realization, and thereby to a single time slot.

The EHU will transmit in a given time slot only when it has harvested enough energy both for processing and transmission in the given time slot, i.e., only when its harvested energy accumulates to a level which is higher than $P_p+P_{\rm EHU}(x_2,v)$, where $v$ is the fading gain in the time slot considered for transmission. Otherwise, the EHU is silent and only harvests energy. When the EHU transmits, it transmits the next untransmitted block of $k$ symbols of its codeword. To this end, each symbol of this block is first multiplied by $\sqrt{P_{EHU}(x_2,v)}$, where $P_{EHU}(x_2,v)$ can be found from Lemma 2, and then the block of $k$ symbols is transmitted over the wireless channel to the ET. The EHU repeats this procedure until it transmits all $N$ blocks of its entire codeword for which it needs $N+B$ time slots.

\emph{Receptions at the ET:} When the ET receives a transmitted block by the EHU, it checks if the power level of the received block is higher than the noise level at the ET or not. If affirmative, the ET places the received block in its data storage, without decoding. Otherwise the received block is discarded.

Now, in $N+B$ time slots, the ET receives the entire codeword transmitted by the EHU. In order for the ET to be able to decode the transmitted codeword, the rate of the transmitted codeword must be equal to or lower than the capacity of the EHU-ET's channel, given by
\begin{align}\label{eqn0b}
C_{EHU-ET}=\frac {1}{2}\sum_{v \in \mathcal{V}} \log \left(1+\frac{v^2 P_{EHU}(x_2,v)}{\sigma_2^2 + x_2^2 \alpha_2}\right)p(v).
\end{align}

Note that the rate of the transmitted codeword is $R_{EHU}$, given by (31). Now, since $R_{EHU} = C_{EHU-ET}$, the ET is able to decode the codeword transmitted by the EHU, by using a joint topicality decoder. Next, since the ET knows the binary sequences corresponding to each group, by decoding the transmitted codeword the ET determines the group to which the transmitted codeword belongs to. As a result, the ET is able to decode the secret message $W$.

In the $N+B$ time slots, the achieved secrecy rate is given by $\lim_{(N+B)\to\infty} \frac{k N}{k (N+B)} R_s \newline = \lim_{(N+B)\to\infty} \frac{N}{N+B} R_s$. It was proven in \cite{eh_zlatanov} that when the EHU is equipped with a battery with an unlimited storage capacity and when C2 in (\ref{cap_eq1}) holds, then $N/(N+B)\to 1$ as $(N+B)\to\infty$. Thereby, the achieved secrecy rate in $N+B$ time slots is given by $\lim_{(N+B)\to\infty} \frac{N}{N+B} R_s = R_s$, which is the actual lower bound of the channel secrecy capacity given by Lemma 2.

\emph{Receptions at the EVE:} EVE receives the transmitted blocks by the EHU and the ET. Similarly to the ET, EVE places the received block in its data storage, without decoding.

In $N+B$ time slots, the EVE also receives the entire codeword transmitted by the EHU. In addition, EVE receives the signal from the ET, comprised of randomly generated symbols (see Lemma 2), which acts as noise to EVE and impairs the ability of EVE to decode the codeword from the EHU. To show that the EVE will not be able to decode the secret message, we use properties of the multiple access channel, resulting from the EHU and the ET transmitting at the same time. The multiple-access capacity region at the EVE formed by the transmission of the EHU and the ET is given by $I(X_1,X_2;Y_3|V,G,F)$. The EVE will be able to decode the EHU's codeword only if one of the following two cases holds, i.e., $R_{EHU} \leq I(X_1,Y_2|X_2,V,G,F)$ when $R_{ET} \leq I(X_2;Y_3|V,G,F)$ or $R_{EHU} \leq I(X_1,Y_2|V,G,F)$ when $R_{ET} \leq I(X_2;Y_3|X_1,V,G,F)$, where $R_{ET}$ is the entropy of the signal generated by the ET and is given by
\begin{align}\label{rateET}
R_{ET} = - \Big[p(x_2) \log_2 p(x_2) + p(x_2) \log_2 p(x_2) \Big].
\end{align}
In (\ref{rateET}), $p(x_2)=1/2$, see Lemma 2. As a result,
\begin{align}\label{rateET_1}
R_{ET} = \log 2 = 1.
\end{align}

\underline{Case 1:} For the EHU's codeword to be decodable at the EVE in this case, \newline
$R_{EHU} \leq I(X_1,Y_2|X_2,V,G,F)$ and $R_{ET} \leq I(X_2;Y_3|V,G,F)$ have to hold. For $I(X_2;Y_3|V,G,F)$ we have
%\begin{footnotesize}
\begin{align} \label{ach_1e1}
&I(X_2;Y_3|V,F,G)=\sum_{v \in \mathcal{V}}\sum_{g \in \mathcal{G}}\sum_{f \in \mathcal{F}} \left[ \int_{-\infty}^{\infty} \frac{1}{2 \sqrt{2 \pi \sigma_{y_3}^2}}\left( e^{-\frac{(y_3 - x_2)^2}{2 \sigma_{y_3}^2}} + e^{-\frac{(y_3 + x_2)^2}{2 \sigma_{y_3}^2}} \right)\right.\nonumber\\
&\left.\times \ln \left(\frac{1}{2\sqrt{2 \pi \sigma_{y_3}^2}} \left( e^{-\frac{(y_3 - x_2)^2}{2 \sigma_{y_3}^2}} + e^{-\frac{(y_3 + x_2)^2}{2 \sigma_{y_3}^2}} \right)\right) d y_3  \right.\nonumber\\
&-\int_{-\infty}^{\infty}\left. \frac{1}{2\sqrt{2 \pi \sigma_{y_3}^2}} e^{-\frac{(z_3 - x_2)^2}{2 \sigma_{y_3}^2}}  \times \ln  e^{-\frac{(z_3 - x_2)^2}{2 \sigma_{y_3}}^2} d z_3 \right] p(v)p(g)p(f),
\end{align}
%\end{footnotesize}
%
where $x_2$ is drawn from (\ref{cap_eq22a_low_1}) or (\ref{cap_low_cond_2}), depending on which case in Lemma 2 holds. According to \cite{e10030200}, we can rewrite (\ref{ach_1e1}) as

%\newpage

\begin{align} \label{ach_1f1}
&I(X_2;Y_3|V,F,G)=\sum_{v \in \mathcal{V}}\sum_{g \in \mathcal{G}}\sum_{f \in \mathcal{F}} \left[\frac{1}{2} \ln \left(2 \pi e \sigma_{y_3}^2\right) + \frac{x_2^2}{f^2 P_{EHU} (x_2, v) + \sigma_3^2}  \right. \nonumber\\
&-\left. \mathcal{I} \left(\frac{x_2}{\sqrt{f^2 P_{EHU} (x_2, v) + \sigma_3^2}}\right) - \frac{1}{2} \ln \left(2 \pi e \sigma_{y_3}^2\right) \right] p(v)p(g)p(f),
\end{align}
%\end{footnotesize}
where $\mathcal{I} (x)$ is defined in (\ref{lem}).
%where $\mathcal{I} (x)$ can be found as
%\begin{align} \label{ach_1g1}
%\mathcal{I} (x) = \frac{2}{\sqrt{2 \pi} x} e^{-x^2/2} \int_{0}^{\infty} e^{-y^2 / 2x} \cosh(y) \ln (\cosh(y)) dy.
%\end{align}
Thereby,
%\begin{footnotesize}
\begin{align}\label{Ret}
&I(X_2;Y_3|V,F,G) \nonumber\\
&= \sum_{v \in \mathcal{V}}\sum_{g \in \mathcal{G}}\sum_{f \in \mathcal{F}} \left[ \frac{x_2^2}{f^2 P_{EHU} (x_2, v) + \sigma_3^2} \right. \left.- \mathcal{I} \left(\frac{x_2}{\sqrt{f^2 P_{EHU} (x_2, v) + \sigma_3^2}}\right) \right] p(v)p(g)p(f).
\end{align}

The maximum value of the term $\frac{x_2^2}{f^2 P_{EHU} (x_2, v) + \sigma_3^2} - \mathcal{I} \left(\frac{x_2}{\sqrt{f^2 P_{EHU} (x_2, v) + \sigma_3^2}}\right)$ in (\ref{Ret}), for any $v,f,$ and $g$ is one \cite{e10030200}. Thereby, in order for $R_{ET} = I(X_2;Y_3|V,F,G) = 1$ to hold, in which case the EVE would decode the secret message, then $\frac{x_2^2}{f^2 P_{EHU} (x_2, v) + \sigma_3^2} - \mathcal{I} \left(\frac{x_2}{\sqrt{f^2 P_{EHU} (x_2, v) + \sigma_3^2}}\right)$ has to be equal to one for all fading realizations $v,f,g$. However, since the fading realizations are outcomes of  continuous RVs, the probability of that happening is zero. Thereby, $I(X_2;Y_3|V,F,G)$ is less than 1 and as a result $R_{ET} > I(X_2;Y_3|V,F,G)$. Consequently, Case 1 can not hold.

\underline{Case 2:} For the EHU's codeword to be decodable at the EVE, then $R_{EHU} \leq I(X_1,Y_2|V,G,F)$ and $R_{ET} \leq I(X_2;Y_3|X_1,V,G,F)$ have to hold. For $I(X_1;Y_3|V,F,G)$ we have
%\begin{footnotesize}
\begin{align} \label{ach_1e}
&I(X_1;Y_3|V,F,G)=\sum_{v \in \mathcal{V}}\sum_{g \in \mathcal{G}}\sum_{f \in \mathcal{F}} \left[ \int_{-\infty}^{\infty} \frac{1}{2 \sqrt{2 \pi \sigma_{y_3}^2}}\left( e^{-\frac{(y_3 - x_2)^2}{2 \sigma_{y_3}^2}} + e^{-\frac{(y_3 + x_2)^2}{2 \sigma_{y_3}^2}} \right)\right.\nonumber\\
&\left.\times \ln \left(\frac{1}{2\sqrt{2 \pi \sigma_{y_3}^2}} \left( e^{-\frac{(y_3 - x_2)^2}{2 \sigma_{y_3}^2}} + e^{-\frac{(y_3 + x_2)^2}{2 \sigma_{y_3}^2}} \right)\right) d y_3  \right.\nonumber\\
&-\int_{-\infty}^{\infty}\left. \frac{1}{2\sqrt{2 \pi \sigma_{3}^2}}\left( e^{-\frac{(z_3 - x_2)^2}{2 \sigma_{3}^2}} + e^{-\frac{(z_3 + x_2)^2}{2 \sigma_{3}^2}} \right)\right.\left. \times \ln \left( e^{-\frac{(z_3 - x_2)^2}{2 \sigma_{3}^2}} + e^{-\frac{(z_3 + x_2)^2}{2 \sigma_{3}^2}} \right) d z_3 \right] p(v)p(g)p(f),
\end{align}
%\end{footnotesize}
%
where $x_2$ is drawn from (\ref{cap_eq22a_low_1}) or (\ref{cap_low_cond_2}), depending on which case in Lemma 2 holds. We can rewrite (\ref{ach_1e}) as
%\begin{footnotesize}
\begin{align} \label{ach_1f}
&I(X_1;Y_3|V,F,G)=\sum_{v \in \mathcal{V}}\sum_{g \in \mathcal{G}}\sum_{f \in \mathcal{F}} \left[\frac{1}{2} \ln \left(2 \pi e \sigma_{y_3}^2\right) + \frac{x_2^2}{f^2 P_{EHU} (x_2, v) + \sigma_3^2} \right. \nonumber\\
& \left. - \mathcal{I} \left(\frac{x_2}{\sqrt{f^2 P_{EHU} (x_2, v) + \sigma_3^2}}\right) - \frac{1}{2} \ln \left(2 \pi e \sigma_{3}^2\right) \right. \left.- \frac{x_2^2}{\sigma_3^2} + \mathcal{I} \left(\frac{x_2}{\sigma_3}\right) \right] p(v)p(g)p(f),
\end{align}
%\end{footnotesize}
where $\mathcal{I} (x)$ is defined in (\ref{lem}).
%where $\mathcal{I} (x)$ can be found as (\ref{ach_1g1}).
%\begin{align} \label{ach_1g}
%\mathcal{I} (x) = \frac{2}{\sqrt{2 \pi} x} e^{-x^2/2} \int_{0}^{\infty} e^{-y^2 / 2x} \cosh(y) \ln (\cosh(y)) dy.
%\end{align}
By rearranging the elements in (\ref{ach_1f}), we can write
\begin{align} \label{ach_1h}
I(X_1;Y_3|V,F,G)&=\sum_{v \in \mathcal{V}}\sum_{g \in \mathcal{G}}\sum_{f \in \mathcal{F}} \left[\frac{1}{2} \ln \left(1+\frac{f^2 P_{EHU}(x_2,v)}{\sigma_3^2}\right) - \Psi \right] p(v)p(g)p(f).
\end{align}
In (\ref{ach_1h}), $\Psi = \frac{x_2^2}{\sigma_3^2} - \mathcal{I} \left(\frac{x_2}{\sigma_3}\right) - \frac{x_2^2}{f^2 P_{EHU} (x_2, v) + \sigma_3^2} + \mathcal{I} \left(\frac{x_2}{\sqrt{f^2 P_{EHU} (x_2, v) + \sigma_3^2}}\right)$. Now, we can lower bound $R_{EHU}$ as
\begin{align}
&R_{EHU} = \frac {1}{2}\sum_{v \in \mathcal{V}} \log \left(1+\frac{v^2 P_{EHU}(x_2,v)}{\sigma_2^2 + x_2^2 \alpha_2}\right)p(v) \nonumber\\
&> \sum_{v \in \mathcal{V}}\sum_{g \in \mathcal{G}}\sum_{f \in \mathcal{F}} \left[\frac{1}{2} \ln \left(2 \pi e \sigma_{y_3}^2\right) \right. + \left.\frac{x_2^2}{f^2 P_{EHU} (x_2, v) + \sigma_3^2} - \mathcal{I} \left(\frac{x_2}{\sqrt{f^2 P_{EHU} (x_2, v) + \sigma_3^2}}\right) \right. \nonumber\\
&\left. - \frac{1}{2} \ln \left(2 \pi e \sigma_{3}^2\right) - \frac{x_2^2}{\sigma_3^2} + \mathcal{I} \left(\frac{x_2}{\sigma_3}\right) \right] p(v)p(g)p(f) \nonumber\\
&=  \sum_{v \in \mathcal{V}}\sum_{g \in \mathcal{G}}\sum_{f \in \mathcal{F}} \left[\frac{1}{2} \ln \left(1+\frac{f^2 P_{EHU}(x_2,v)}{\sigma_3^2}\right) - \Psi \right] p(v)p(g)p(f)= I(X_1,Y_2|V,G,F).
\end{align}
Thereby, $R_{EHU} > I(X_1,Y_2|V,G,F)$. As a result, Case 2 can not hold either.

As Case 1 and Case 2 show, the EHU's codeword is outside the multiple-access capacity region at the EVE, and thereby the EVE will not be able to decode the EHU's secret message.

\vspace{-3mm}
\section{Numerical results}
In this section, we illustrate numerical examples of the upper bound on the secrecy capacity as well as the derived achievable secrecy rate, and compare it with the achievable secrecy rates of a chosen benchmark scheme. To this end, we first outline the system parameters, then we introduce the benchmark scheme, and finally we provide the numerical results.
%\vspace{-4mm}
\vspace{-3mm}
\subsection{System Parameters}
We use the standard path loss model given by
\begin{align}\label{num_1}
\Omega_j=\left(\frac{c}{f_c 4 \pi}\right)^2 d^{-\gamma}, j \in \{V, F, G\} 
\end{align}
in order to compute the average power of the channel fading gains, $V$, $F$, and $G$, where $c$ denotes the speed of light, $f_c$ is the carrier frequency, $d$ is the length of the considered link (i.e., the length of the EHU-ET link for $\Omega_V$ , the length of the EHU-EVE link for $\Omega_F$ , and the length of the ET-E link for $\Omega_G$), and $\gamma$ is the path loss exponent. We assume that $\gamma=3$. For the carrier frequency, we adopt a value which is commonly associated with low power networks, and we set $f_c = 2.4$ GHz. In addition, we assume a bandwidth of $B = 100$ kHz. To account for the energy losses during the energy harvesting process, the energy harvesting efficiency coefficient $\eta$ is assumed to be equal to 0.8. Throughout this section, we assume Rayleigh fading with average power $\Omega_V$, $\Omega_F$, and $\Omega_G$, respectively, given by (42). To provide practical results, we chose an achievable value for the self-interference suppression factor, and we set $\alpha_2$ to $-100$ dB. As a result, the ET needs to suppress $100$ dB of the self-interference, which is possible in practice [24]. As the EHU does not suppress the self-interference, we chose higher values for the self-interference amplification factor at the EHU and we set $\alpha_1$ to $-40$ dB and $q_1$ to $0$ dB. The noise power to $-90$ dBm.
%The system parameters are summarized in Table I.

%\begin{table}[]
%\centering
%\caption{Simulation parameters}
%\label{my-label}
%\begin{tabular}{ll}
%\hline
%Parameter                  & Value                      \\ \hline
%Speed of light $c$	       & 299 792 458 m / s          \\ \hline
%Carrier frequency $f_c$	   & 2.4 GHz					\\ \hline
%Bandwidth $B$              & 100 kHz                    \\ \hline
%Noise power $\sigma^2$     & -90 dBm            \\ \hline
%%Noise power at the ET $\sigma_3^2$     & -95 dBm            \\ \hline
%Self-interference amplification factors $\alpha_1$, $\alpha_2$     &-100 dBm, -40 dBm            \\ \hline
%EH efficiency $\eta$       & 0.8                        \\ \hline
%Path loss exponent $\gamma$& 3                          \\ \hline
%EHU-ET distance $d_{EHU-ET}$       		   & 10 m 		        \\ \hline
%EHU-EVE distance $d_{EHU-EVE}$       		   & 11 m 		        \\ \hline
%ET-EVE distance $d_{ET-EVE}$       		       & 12 m 		        \\ \hline
%Processing cost $P_p$      & -40 dBm                      \\ \hline
%%ET transmit power $P_{ET}$ & 1 W $\sim$ 35 dBm                 \\ \hline
%\end{tabular}
%\end{table}

\vspace{-3mm}
\subsection{Benchmark Scheme}
Since to the best of the authors' knowledge there are no available communication schemes in the literature for the considered FD system model, we use the HD counterpart as a benchmark scheme, which is outlined in the following.

%\textit{Benchmark Scheme:}
Time is divided into time slots with duration $T$. A single time slot coincides with one fading realization. A portion of each time slot, denoted by $T-t$, is used for energy transmission by the ET and for energy harvesting by the EHU and the rest of the time slot, $t$, is used for information transmission by the EHU, during which the ET is silent. Hence the EHU and the ET both operate in the HD mode. The EHU and the ET are assumed to have full CSI of the EHU-ET channel. Since in this case the ET stops transmitting during the information transmission by the EHU, an interference signal is not present at the EVE. 
The secrecy rate is thus given by
\begin{align}\label{num_1}
R_s=&\max \left( 0 , \max_t t \left(\frac{1}{2} \sum_{v\in\mathcal{V}} \log \left(1+\frac{v^2 P_{EHU}(v)}{\sigma_1^2}\right)p(v) \right.\right. \nonumber\\
&\left.\left.-\sum_{v\in\mathcal{V}}\sum_{f\in\mathcal{F}}\log \left(1+\frac{f^2 P_{EHU}(v)}{\sigma_3^2}\right)p(v)p(f) \right) \right),
\end{align}
where $P_{EHU}(v)$ can be found as the root of the following equation
\begin{align}\label{num_1a}
&\frac{v^2}{\sigma_1^2}-\left(1+\frac{v^2 P_{EHU}(v)}{\sigma_1^2}\right) \sum_{f\in\mathcal{F}} \frac{f^2}{f^2 P_{EHU}(v)+\sigma_3^2} p(f)
=\left(1+\frac{v^2 P_{EHU}(v)}{\sigma_1^2}\right) \lambda_2
\end{align}
and $\lambda_2$ is chosen such that
\begin{align}\label{num_1b}
&t \left(\sum_{v\in\mathcal{V}} P_{EHU} (v) p(v) + P_p\right) = (T-t) \eta P_{ET} \Omega_V
\end{align}
holds.

\vspace{-3mm}
\subsection{Numerical Examples}
\begin{figure*}[tbp]
\centering
\includegraphics[width=6.1in,height=3.0in]{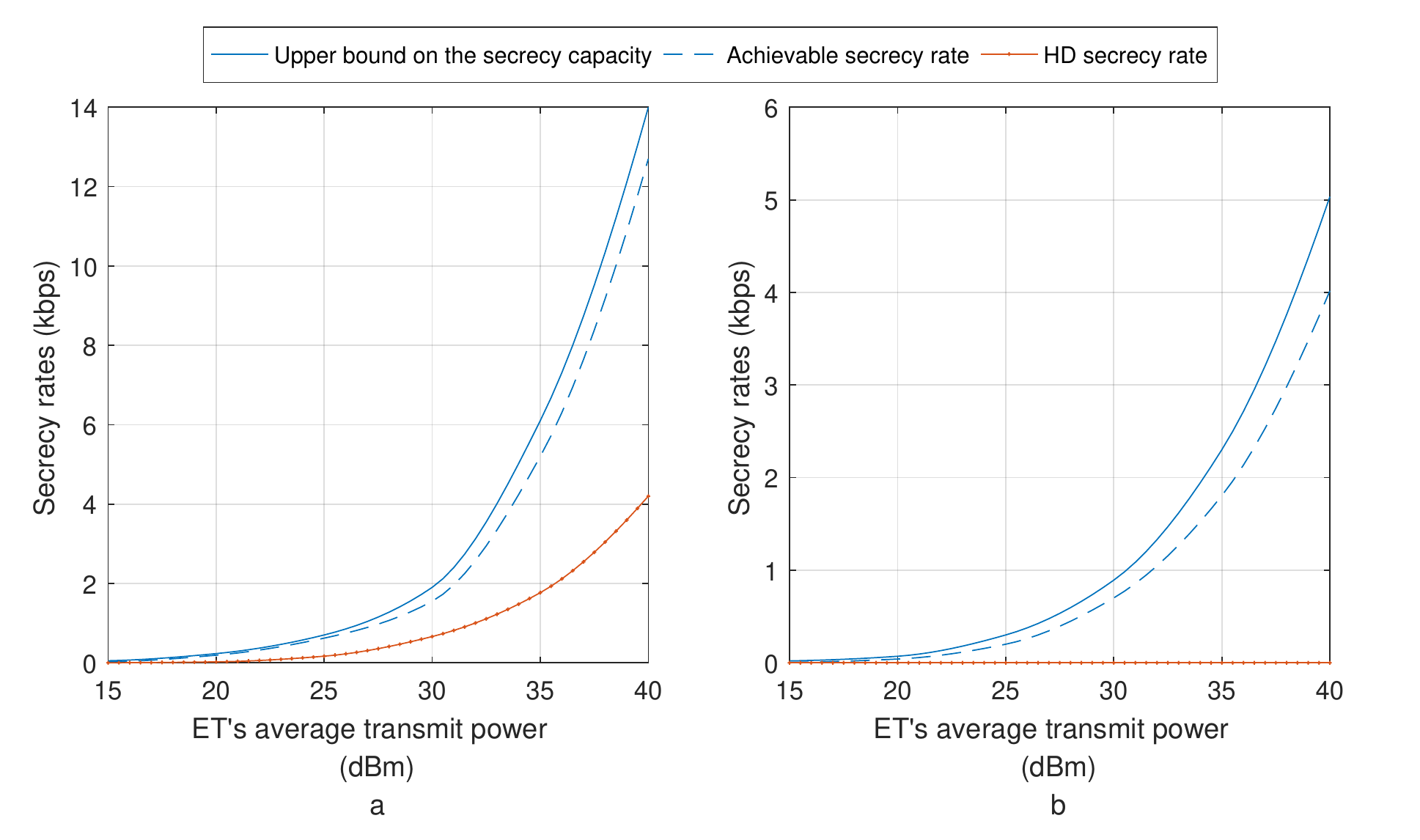}
\caption{Upper and lower bounds on the secrecy capacity. Fig. 1a corresponds to the case when the ET is closer to the EHU than the EVE, i.e., $d_{EHU-ET}=10m$ and $d_{EHU-EVE} = 12m$, whilst Fig. 1b corresponds to the case when the EVE is closer to the EHU than the ET, i.e., $d_{EHU-ET}=10m$ and $d_{EHU-EVE}=9m$. }
\label{num1}
\end{figure*}

The upper and lower bounds on the secrecy capacity are illustrated on Fig.~\ref{num1}, and are evaluated against the benchmark scheme. From Fig.~\ref{num1} we notice that the FD scheme outperforms the HD scheme, which is a result of two factors. Firstly, energy recycling is impossible when the EHU operates as an HD node. Secondly, in HD, the ET stops acting like a jammer and an interference signal is not present at the EVE. Note that since physical layer security relies on the channel quality between the nodes in the network, distance plays a crucial role. To demonstrate this, we consider different cases for the distances between the nodes. In the case when the ET is closer to the EHU than the EVE, the EHU-ET channel is better, on average, than the EHU-EVE channel. For this case, the secrecy rate of the HD benchmark scheme is non-zero, but it is smaller than the  derived achievable secrecy rate of the FD scheme, as it can be seen in Fig. 1a. When the EVE is closer to the EHU than the ET, the EHU-ET channel is worse, on average, than the EHU-EVE channel. In this case the secrecy rate of the HD benchmark scheme is zero, whereas the derived FD secrecy rate is positive, see Fig. 1b. Thereby, as a result of the interference signal generated by the ET, in these cases the proposed FD scheme offers positive secrecy rates even when the EVE is closer to the EHU than the legitimate receiver, which is impossible to achieve by employing the HD scheme. As our numerical results show,  the difference between the upper bound on the secrecy capacity and the achievable rate is smaller than 1 dB. Thereby, the derived upper bound is tight in the sense that it shows that the secrecy capacity and the achievable rate differ only 1 dB, which for practical purposes is not a large gap.

\begin{figure}[tbp]
\centering
\includegraphics[width=4.1in,height=3.0in]{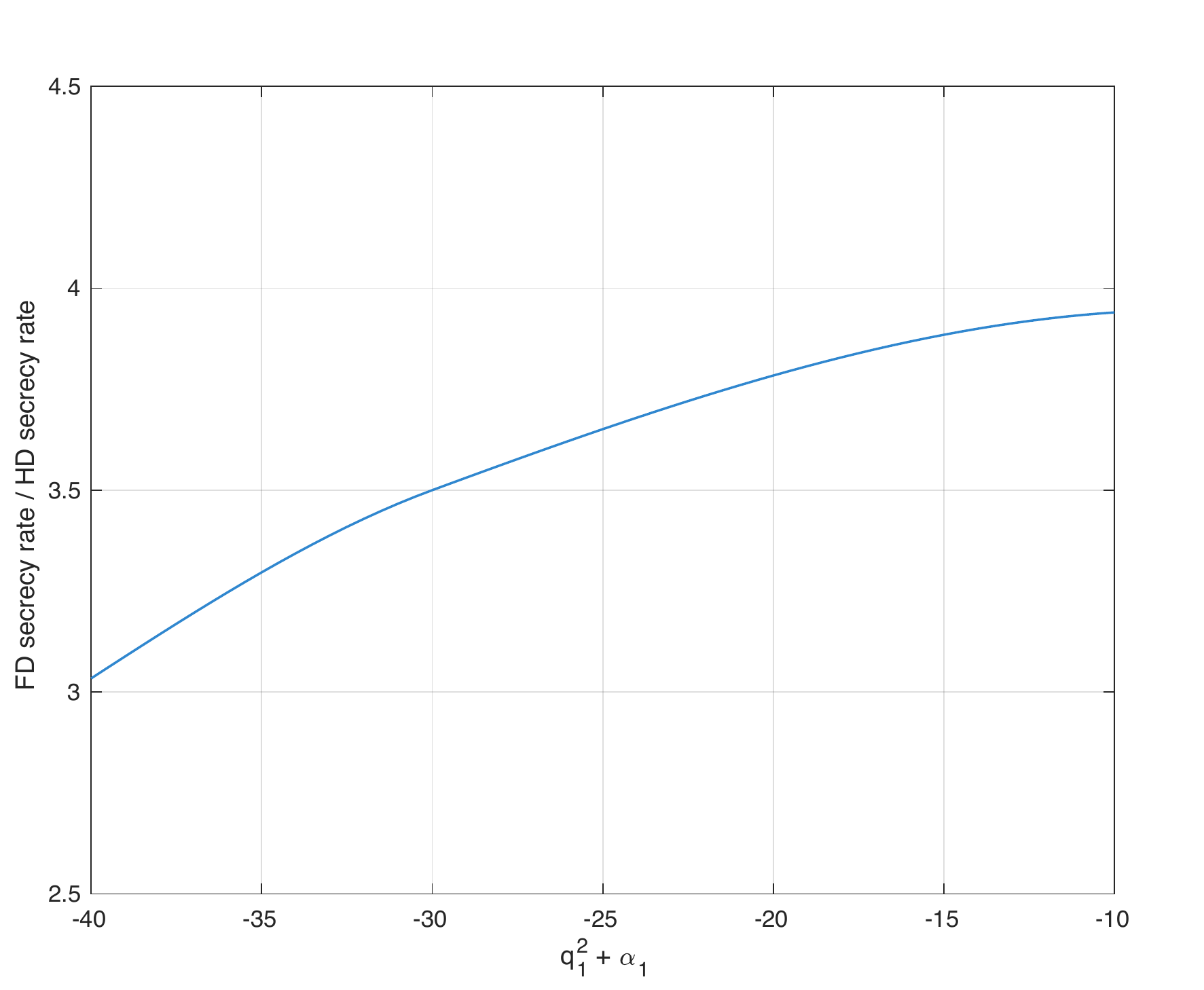}
\caption{FD secrecy rate / HD secrecy rate ratio as a function of the self-interference at the EHU. Fig. 2 corresponds to the case when the ET is closer to the EHU than the EVE, i.e., $d_{EHU-ET}=10m$ and $d_{EHU-EVE} = 12m$. }
\label{num2}
\end{figure}

\begin{figure}[tbp]
\centering
\includegraphics[width=4.1in,height=3.0in]{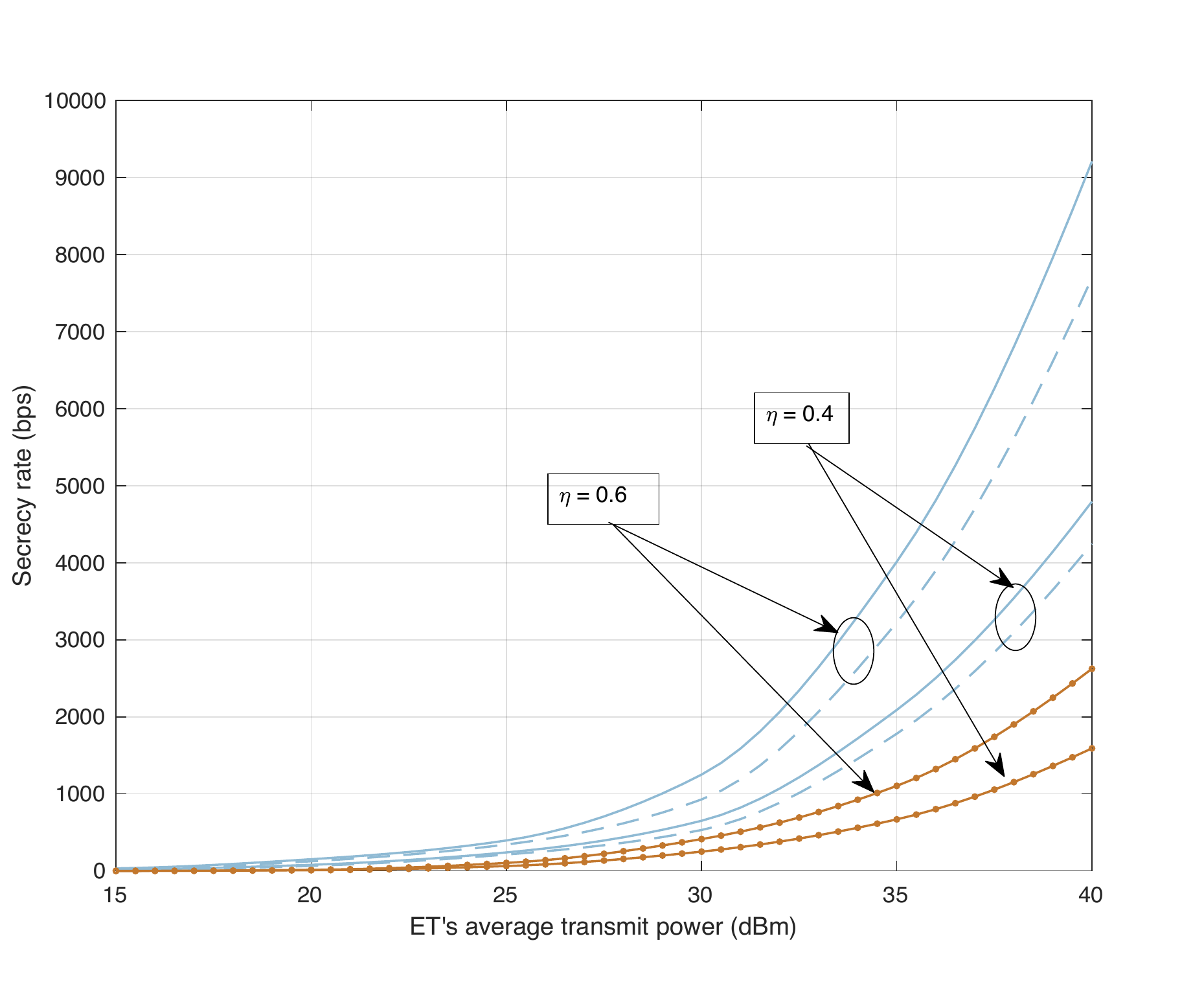}
\caption{Upper and lower bounds on the secrecy capacity, for different values of the energy harvesting efficiency coefficient $\eta$. Fig. 3 corresponds to the case when the ET is closer to the EHU than the EVE, i.e., $d_{EHU-ET}=10m$ and $d_{EHU-EVE} = 12m$. }
\label{num3}
\end{figure}

In Figs.~\ref{num2}-~\ref{num3}, we show the impact of the self-recycled energy on the achieved secrecy rates. In particular, we show the ratio between the FD secrecy rate and the HD secrecy rate as a function of the self-interference at the EHU, given by $q_1^2+\alpha_1$. As the self-interference at the EHU increases, so does the FD secrecy rate. Meanwhile, the HD secrecy rate is constant, as the HD mode does not result in self-interference. Thereby, higher self-interference at the EHU leads to higher secrecy rate. The secrecy rates are also illustrated on Fig.~\ref{num3}, for different values of the energy harvesting efficiency coefficient $\eta$. As the energy harvesting efficiency coefficient represents the amount of energy that is converted from harvested to operational energy (i.e., energy used for transmission and processing), lower values for $\eta$ yield lower secrecy rates, both for FD and HD.  However, the FD scheme again outperforms the HD scheme as a result of the previously discussed factors.

\vspace{-3mm}
\section{Conclusion}
In this paper, we have studied the secrecy capacity of a FD wirelessly powered communication system consisting of an EHU and an ET in the presence of a passive EVE. We have shown that the ET's transmit signal can act as interference against the EVE. We have derived an upper bound on the secrecy capacity and, furthermore, an achievable secrecy rate that can be achieved with relative low complexity. It has been shown that the proposed scheme achieves higher secrecy rates compared to the conventional HD-based schemes, even in the case when the channel between the EHU and the ET is worse, on average, than the channel between the EHU and EVE. In future works, it would be interesting to consider an arbitrary number of EHUs and EVEs in the network.

\vspace{-5mm}
\appendices
\section{Converse}\label{Con}
In order for us to claim that the result in (\ref{cap_eq1}) is indeed an upper bound on the secrecy capacity of the considered channel, we provide the following converse.

Let $W$ be the confidential message that the EHU wants to transmit to the ET and which EVE wants to intercept. Let this message be uniformly selected at random from the message set $\{1,\,2,\,...,2^{nR_s}\}$, where $n\to\infty$ is the number of channel uses that will be used for transmitting $W$ from the EHU to the ET, and  $R_s$ denotes the data rate of   message $W$. We assume a priori knowledge of the CSI of the EHU-ET channel, i.e., $V_i$ is known for $i=1....n$ before the start of the communication at all three nodes. In addition, the EHU-EVE and the ET-EVE channels, given by $G_i$ and $F_i$, respectively, are only known by EVE for $i=1....n$.

We have the following limits for the mutual information between the EHU and EVE
\begin{align}\label{eq_1aa}
&I(W;Y_3^n|V^n,G^n,F^n)=H(W|V^n,G^n,F^n)-H(W|Y_3^n, V^n,G^n,F^n)\nonumber\\
&\stackrel{(a)}{\leq} H(W|V^n)-H(W|Y_3^n, V^n,G^n,F^n)\leq n\epsilon,
\end{align}
where $(a)$ follows since conditioning reduces entropy and $\epsilon$ is a positive number. On the other hand, we have the following limit due to Fano's inequality \cite{cover2012elements}
\begin{align}\label{eq_1b}
 H(W|Y_2^n,V^n)\leq P_e n R_s+1,
\end{align}
where $P_e$ is the average probability of error of the message $W$ and $R_s$ is the secrecy rate.

Now, for the secrecy rate, $R_s$, we have the following limit
\begin{align}\label{eq_1}
& nR_s\leq H(W|V^n) \stackrel{(a)} \leq H(W|Y_3^n, V^n, G^n, F^n) + n\epsilon =H(W|Y_3^n, V^n, G^n, F^n)+n\epsilon\nonumber\\
& +H(W|V^n)-H(W|V^n)+H(W|V^n, Y_2^n,X_2^n)-H(W|V^n, Y_2^n,X_2^n)\nonumber\\
& \stackrel{(b)}\leq
H(W|Y_3^n, V^n, G^n, F^n)+n\epsilon\nonumber\\
& +H(W|V^n)-H(W|V^n, G^n, F^n)+H(W|V^n, Y_2^n,X_2^n)-H(W|V^n, Y_2^n,X_2^n)
\nonumber\\
&\stackrel{(c)}=I(W;Y_2^n,X_2^n|V^n)-I(W;Y_3^n|V^n, G^n, F^n)+H(W|V^n, Y_2^n,X_2^n)+ n\epsilon \nonumber\\
&\stackrel{(d)}\leq I(W;Y_2^n,X_2^n|V^n)-I(W;Y_3^n|V^n, G^n, F^n)+H(W|V^n, Y_2^n)+ n\epsilon \nonumber\\
&\stackrel{(e)}\leq I(W;Y_2^n,X_2^n|V^n)-I(W;Y_3^n|V^n, G^n, F^n)+P_e n R_s + 1 + n\epsilon
\end{align}
where $(a)$ follows from (\ref{eq_1aa}), $(b)$ follows from the fact that conditioning reduces entropy, $(c)$ is obtained by exploiting $I(W;Y_2^n,X_2^n|V^n)=H(W|V^n)-H(W|V^n, Y_2^n,X_2^n)$ and $I(W;Y_3^n|V^n, G^n, F^n)=H(W|V^n, G^n, F^n)-H(W|Y_3^n, V^n, G^n, F^n)$, $(d)$ results from the fact that conditioning reduces entropy, and $(e)$ follows by  Fano's inequality given by (\ref{eq_1b}).
Dividing both sides of (\ref{eq_1}) by $n$, we have
\begin{align}\label{eq_3}
R_s\leq \frac{1}{n}I(W; Y_2^n,X_2^n|V^n)-\frac{1}{n}I(W;Y_3^n|V^n, G^n, F^n)+P_e  R_s +\frac{1}{n} + \epsilon.
\end{align}
Assuming that  $P_e\to 0$   and $\epsilon \to 0$ as $n\to \infty$, which means that we assume a zero-error probability at the ET and zero mutual information between the EHU and EVE,
 (\ref{eq_3}) for $n \to \infty$ can be written as
\begin{align}\label{eq_4}
R_s\leq \frac{1}{n} I(W; Y_2^n,X_2^n|V^n)-\frac{1}{n}I(W;Y_3^n|V^n, G^n, F^n).
\end{align}
We represent the first element of the right hand side of (\ref{eq_4}) as
\begin{align}\label{eq_5}
I(W;Y_2^n,X_2^n|V^n)&=I(W;Y_2^n|X_2^n,V^n)+I(W;X_2^n|V^n).
\end{align}
Now, since the transmitted message $W$ is uniformly drawn from the message set at the EHU and since the ET does not know which message the EHU transmits, the following holds
\begin{align}\label{eq_5a}
I(W;X_2^n|V^n)=0.
\end{align}
Inserting (\ref{eq_5a}) into (\ref{eq_5}), we have
\begin{align}\label{eq_5aa}
I(W;Y_2^n,X_2^n|V^n)&=I(W;Y_2^n|X_2^n,V^n).
\end{align}
Inserting (\ref{eq_5aa}) into (\ref{eq_4}), we have
\begin{align}\label{eq_6}
& R_s\leq \frac{1}{n} I(W; Y_2^n|X_2^n,V^n)-\frac{1}{n}I(W;Y_3^n|V^n, G^n, F^n) \nonumber\\
&\stackrel{(a)}{\leq}\sum_{i=1}^n \Big(I(W;Y_{2i}|Y_2^{i-1},X_2^n,V^n)- I(W;Y_{3i}|Y_3^{i-1}, V^n, G^n, F^n)\Big)\nonumber\\
%&\textbf{OD OVDEKA PA NADOLE REVIDIRAJ}
%\nonumber\\
%&\stackrel{(b)}{\leq} \frac{1}{n}\sum_{i=1}^n \left(I(X_1^n;Y_{2i}|Y_2^{i-1},X_2^n,V^n)-I(X_1^n;Y_{3i}|Y_3^{i-1} V^n, G^n, F^n)\right)\nonumber\\
& =\frac{1}{n} \sum_{i=1}^n \Big(H(Y_{2i}|Y_2^{i-1},X_2^n,V^n)-H(Y_{2i}|Y_2^{i-1},X_2^n,V^n,W)  \nonumber\\
&-H(Y_{3i}|Y_3^{i-1} V^n, G^n, F^n)+H(Y_{3i}|Y_3^{i-1}, V^n, G^n, F^n,W)\Big) \nonumber\\
&\stackrel{(b)}{\leq} \frac{1}{n}\sum_{i=1}^n \Big(H(Y_{2i}|Y_2^{i-1},X_2^n,V^n)-H(Y_{2i}|Y_2^{i-1},X_2^n,V^n,W,X_{1i}) \nonumber\\
&-H(Y_{3i}|Y_3^{i-1}, V^n, G^n, F^n)+H(Y_{3i}|Y_3^{i-1}, V^n, G^n, F^n,W)\Big).
\end{align}
where $(a)$ follows from the fact that the entropy between a collection of random
variables is less than or equal to the sum of their individual entropies and $(b)$ results from the fact that conditioning reduces entropy.
On the other hand, because of the memoryless channel assumption, $Y_{3i}$ is independent of $Y_{3}^{i-1}$, therefore, we can write
\begin{align}\label{eq_6a}
  &  H(Y_{3i}|Y_3^{i-1}, V^n, G^n, F^n,W) = H(Y_{3i}| V^n, G^n, F^n,W) \nonumber\\
   & \stackrel{(a)}{=} H(Y_{3i}, V^n, G^n, F^n,W)-H(V^n, G^n, F^n,W) \stackrel{(b)}{\leq} H(Y_{3i}, V^n, G^n, F^n,W, X_1^n)-H(V^n, G^n, F^n,W)\nonumber \\
   & \stackrel{(c)}{=} H(Y_{3i} | V^n, G^n, F^n,W, X_1^n)+H(V^n, G^n, F^n,W, X_1^n) -H(V^n, G^n, F^n,W) \nonumber\\
   & \stackrel{(d)}{=} H(Y_{3i} | V^n, G^n, F^n,W, X_1^n)+H(X_1^n|V^n, G^n, F^n, W) +H(V^n, G^n, F^n,W)-H(V^n, G^n, F^n,W) \nonumber\\
   & = H(Y_{3i} | V^n, G^n, F^n,W, X_1^n)+H(X_1^n|V^n, G^n, F^n, W) \stackrel{(e)}{=} H(Y_{3i} | V^n, G^n, F^n,W, X_1^n) \nonumber\\
   & \stackrel{(f)}{\leq} H(Y_{3i} | V^n, G^n, F^n, X_1^n),
\end{align}
where $(a)$ follows from the chain rule for joint entropy, $(b)$ follows from the properties of joint entropy, $(c)$ and $(d)$ follow from the chain rule for joint entropy, $(e)$ follows from the fact that $H(X_1^n|W,V^n, G^n, F^n)=0$ because of the deterministic mapping $W\rightarrow X_1^n$, and $(f)$ follows from the fact that conditioning reduces entropy.

By inserting (\ref{eq_6a}) into (\ref{eq_6}), we obtain
\begin{align} \label{eq_6b}
&R_s\leq \frac{1}{n}\sum_{i=1}^n \Big(H(Y_{2i}|Y_2^{i-1},X_2^n,V^n)-H(Y_{2i}|Y_2^{i-1},X_2^n,V^n,X_{1i},W) \nonumber\\
&-H(Y_{3i}|Y_3^{i-1},V^n, G^n, F^n)+H(Y_{3i}|V^n, G^n, F^n,X_1^n)\Big)\nonumber\\
&\stackrel{(a)}{=} \frac{1}{n}\sum_{i=1}^n \Big(H(Y_{2i}|X_{2i},V_i)-H(Y_{2i}|X_{2i},V_i,X_{1i},W)-\left(H(Y_{3i}|V_i, G_i, F_i)-H(Y_{3i}|V_i, G_i, F_i,X_{1i})\right)\Big) \nonumber\\
&\stackrel{(b)}{=} \frac{1}{n}\sum_{i=1}^n \Big(H(Y_{2i}|X_{2i},V_i)-H(Y_{2i}|X_{2i},V_i,X_{1i})-\left(H(Y_{3i}|V_i, G_i, F_i)-H(Y_{3i}|V_i, G_i, F_i,X_{1i})\right)\Big)
\end{align}
where $(a)$ follows from the fact that due to the memoryless channel assumption, $Y_{2i}$ is independent of all elements in the vectors $X_2^n$, $V^n$, and $X_1^n$ except the elements  $X_{2i}$, $V_i$, and $X_{1i}$, respectively, and of $Y_2^{i-1}$, and thereby
$H(Y_{2i}|Y_2^{i-1},X_2^n,V^n)=H(Y_{2i}|X_{2i},V_i)$, and \\
$H(Y_{2i}|Y_2^{i-1},X_2^n,V^n,X_{1i},W)=H(Y_{2i}|X_{2i},V_i,X_{1i},W)$. Similarly, $Y_{3i}$ is independent of all the elements of the vector $X_1^n$ except $X_{1i}$, of all the elements of the vector $V^n$ except $V_i$, of all the elements of the vector $G^n$ except $G_i$, of all the elements of the vector $F^n$ except $F_i$ and of $Y_3^{i-1}$, and thereby $H(Y_{3i}|Y_3^{i-1},V^n, G^n, F^n)=H(Y_{3i}|V_i, G_i, F_i)$ and $H(Y_{3i}|V^n, G^n, F^n,X_1^n)=H(Y_{3i}|V_i, G_i, F_i,X_{1i})$. In continuation, $(b)$ follows from the fact that given $X_{2i}$, $V_i$, and $X_{1i}$, $Y_{2i}$ is conditionally independent of the message $W$ as it can be seen from (\ref{eq_5a1}), and thereby $H(Y_{2i}|X_{2i},V_i,X_{1i},W)=H(Y_{2i}|X_{2i},V_i,X_{1i})$.
Now,  we can write (\ref{eq_6b}) as
\begin{align}\label{eq_7}
R_s&\leq \frac{1}{n} \sum_{i=1}^n \Big(I(X_{1i};Y_{2i}|X_{2i},V_i)-I(X_{1i};Y_{3i}|V_i, G_i, F_i) \Big)=I(X_1;Y_{2}|X_{2},V) - I(X_1;Y_3|V, G, F).
\end{align}
Therefore, an upper bound on the secrecy capacity is given by (\ref{eq_7}) when no additional constraints on $X_1$ and $X_2$ exist and it is achieved by maximizing over all possible probability distributions $p(x_1,x_2|v)$, or equivalently by $\{p(x_1|x_2,v),p(x_2|v)\}$. In our case, we impose a further constraint on $X_2$ which limits the ET's average output power to $P_{ET}$, which is expressed by C1 in (\ref{cap_eq1}). Moreover, the second constraint, expressed by C2 in (\ref{cap_eq1}), concerns $X_1$ and it limits the average transmit power of the EHU to be less than the maximum average harvested power minus the processing cost $P_p$. Constraints C3 and C4 in (\ref{cap_eq1}) come from the definitions of probability distributions. Hence, the capacity is upper bounded by (\ref{cap_eq1}). This proves the converse.
\section{Proof of Theorem 2}

Since the EHU-ET channel is an AWGN channel with channel gain $v$ and AWGN with variance $\sigma_2^2 + x_2^2 \alpha_2$, $I(X_{1};Y_{2}|X_{2}=x_2, V=v)=\frac {1}{2} \log \left(1+\frac{v^2 P_{EHU}(x_2,v)}{\sigma_2^2 + x_2^2 \alpha_2}\right)$. In addition, since $Q_1$ and $X_1$ are zero-mean Gaussian RVs, the left-hand side of constraint C2 in (\ref{cap_eq1}) can be transformed into
\begin{flalign} \label{appB_eq0}
&\int_{x_1} \sum_{x_2 \in \mathcal{X}_2} \sum_{v \in \mathcal{V}} (x_1^2 + P_p) p(x_1|x_2,v)p(x_2|v)p(v)dx_1 =\sum_{x_2 \in \mathcal{X}_2} \sum_{v \in \mathcal{V}} P_{EHU}(x_2,v) p(x_2|v)p(v) +P_p.
\end{flalign}
where we have used $\int_{x_1} \sum_{x_2 \in \mathcal{X}_2} \sum_{v \in \mathcal{V}} x_1^2 p(x_1|x_2,v)p(x_2|v)p(v)dx_1\\
=\sum_{x_2 \in \mathcal{X}_2} \sum_{v \in \mathcal{V}} P_{EHU}(x_2,v) p(x_2|v)p(v)$. Whereas, the right-hand side of C2 in (\ref{cap_eq1}) can be rewritten as
\begin{flalign}\label{appB_eq0aa}
&\int_{x_1} \sum_{x_2 \in \mathcal{X}_2} \sum_{v \in \mathcal{V}} E_{\rm in} p(x_1|x_2,v)p(x_2|v)p(v)d x_1  \nonumber\\
&=\int_{q_1} \int_{x_1} \sum_{x_2 \in \mathcal{X}_2} \sum_{v \in \mathcal{V}} \eta(e x_2 + \bar q_{1} x_{1}+q_{1} x_{1})^2 p(x_1|x_2,v)p(x_2|v)p(v)p(q_1)d x_1 d q_1  \nonumber\\
&=\sum_{x_2 \in \mathcal{X}_2} \sum_{v \in \mathcal{V}} \eta v^2 x_{2}^2 p(x_2|v)p(v) + \int_{x_1}\sum_{x_2 \in \mathcal{X}_2} \sum_{v \in \mathcal{V}} \eta \bar q_{1}^2 x_{1}^2 p(x_1|x_2,v)p(x_2|v)p(v)d x_1  \nonumber\\
&+\int_{q_1} \int_{x_1}\sum_{x_2 \in \mathcal{X}_2} \sum_{v \in \mathcal{V}} \eta q_{1}^2 x_{1}^2 p(x_1|x_2,v)p(x_2|v)p(v)p(g_1)d x_1 d g_1 \nonumber\\
&= \sum_{x_2 \in \mathcal{X}_2} \sum_{v \in \mathcal{V}} \eta e^2 x_{2}^2 p(x_2|v)p(v) +\eta \bar q_{1}^2 \sum_{x_2 \in \mathcal{X}_2} \sum_{v \in \mathcal{V}} P_{EHU}(x_2,v) p(x_2|v)p(v)\nonumber\\
&+ \eta \alpha_1 \sum_{x_2 \in \mathcal{X}_2} \sum_{v \in \mathcal{V}} P_{EHU}(x_2,v)  p(x_2|v)p(v),
\end{flalign}
where $q_1$ represents the realizations of the random variable $Q_1$. Combining (\ref{appB_eq0}) and (\ref{appB_eq0aa}) transforms (\ref{cap_eq1}) into

\begin{align} \label{appB_eq0a}
&\max_{P_{EHU}(x_2,v), p(x_2|v)} \sum_{x_2 \in \mathcal{X_2}} \sum_{v \in \mathcal{V}}{ \frac{1}{2} \log \left(1+\frac{v^2 P_{EHU}(x_2,v)}{\sigma_2^2 + x_2^2 \alpha_2} \right) p(x_2|v)p(v)}\nonumber\\
&\qquad\qquad- \sum_{v \in \mathcal{V}}\sum_{g \in \mathcal{G}}\sum_{f \in \mathcal{F}}  I(X_{1};Y_{3}| V=v, G=g, F=f)  p(v)p(g)p(f) \nonumber\\
&{\rm{Subject\;\;  to}}  \nonumber\\
&\qquad\qquad{\rm C1:} \sum_{x_2 \in \mathcal{X}_2} \sum_{v \in \mathcal{V}} x_2^2 p(x_2|v)p(v) \leq P_{ET} \nonumber\\
&\qquad\qquad{\rm C2:} \sum_{x_2 \in \mathcal{X}_2} \sum_{v \in \mathcal{V}} P_{EHU}(x_2,v)p(x_2|v)p(v) + P_p \leq \nonumber\\
&\qquad\qquad\qquad  \sum_{x_2 \in \mathcal{X}_2} \sum_{v \in \mathcal{V}} \eta v^2 x_2^2 p(x_2|v)p(v) \nonumber\\
&\qquad\qquad\qquad + \eta(\bar {q_1}^2 +\alpha_1)\sum_{x_2 \in \mathcal{X}_2} \sum_{v \in \mathcal{V}} P_{EHU}(x_2,v) p(x_2|v)p(v)\nonumber\\
&\qquad\qquad{\rm C3:} \sum_{x_2 \in \mathcal{X}_2}{p(x_2|v)} = 1 \nonumber\\.
&\qquad\qquad{\rm C4:} P_{EHU}(x_2,v) \geq 0 .
\end{align}

Now, since the log function and the mutual information are both concave functions \cite{cover} with respect to the optimization variables, their difference, as given in the objective function of (\ref{appB_eq0a}) is in general neither concave nor convex. Therefore, the optimization problem in (\ref{appB_eq0a}) may not be convex so a given solution can either be a local maximum or a global maximum. However, since we are interested in finding an upper bound on the secrecy capacity, we can still apply the Lagrange duality method due to the fact that the dual function of a maximization optimization problem yields an upper bound on the optimal solution, see \cite{Boyd_CO}.  Thereby, we write the Lagrangian of (\ref{appB_eq0a}) as
\begin{flalign} \label{appB_eq0b}
&\mathcal{L}=\sum_{x_2 \in \mathcal{X_2}} \sum_{v \in \mathcal{V}}{\frac {1}{2} \log \left(1+\frac{v^2 P_{EHU}(x_2,v)}{\sigma_2^2 + x_2^2 \alpha_2}\right)p(x_2|v)p(v)}\nonumber\\
&-\sum_{v \in \mathcal{V}}\sum_{g \in \mathcal{G}}\sum_{f \in \mathcal{F}} I(X_{1};Y_{3}|V=v, G=g, F=f) p(v)p(g)p(f) \nonumber\\
&-\lambda_1 \left(\sum_{x_2 \in \mathcal{X}_2} \sum_{v \in \mathcal{V}} x_2^2 p(x_2|v)p(v)- P_{ET}\right)-\lambda_2 \left((1-\eta(\bar {q_1}^2 +\alpha_1))\sum_{x_2 \in \mathcal{X}_2} \sum_{v \in \mathcal{V}} P_{EHU}(x_2,v) p(x_2|v)p(v) \right. \nonumber\\
&\left. \left. + P_p -\sum_{x_2 \in \mathcal{X}_2} \sum_{v \in \mathcal{V}} \eta v^2 x_2^2 p(x_2|v)p(v)\right) \right.-\mu_1 \left(\sum_{x_2 \in \mathcal{X_2}}{p(x_2|v)}-1\right) - \mu_2 P_{EHU}.  \nonumber\\
\end{flalign}
In (\ref{appB_eq0a}), we assume that $0<\eta(\bar {g_1}^2 +\alpha_1)<1$, since $\eta(\bar {g_1}^2 +\alpha_1)\geq1$ would practically imply that the EHU recycles the same or even a larger amount of energy than what has been transmitted by the EHU, which is not possible in reality. In (\ref{appB_eq0b}), $\lambda_1$, $\lambda_2$, $\mu_1$, and $\mu_2$ are the Lagrangian multipliers associated with C1, C2, C3, and C4 in (\ref{cap_eq1}), respectively. Differentiating (\ref{appB_eq0b}) with respect to the optimization variables, we obtain
\begin{align}
&\frac{\partial \mathcal{L}}{\partial P_{EHU}(x_2,v)}=\frac{\frac{v^2}{\sigma_2^2 + x_2^2 \alpha_2}}{1+\frac{v^2 P_{EHU}(x_2,v)}{\sigma_2^2 + x_2^2 \alpha_2}} -\lambda_2 (1-\eta(\bar {g_1}^2 +\alpha_1)) -\mu_2 \nonumber\\
&-\frac{\partial}{\partial P_{EHU}(x_2,v)} \left( \sum_{g \in \mathcal{G}}\sum_{f \in \mathcal{F}} I(X_{1};Y_{3}|V=v, G=g, F=f) p(g)p(f)\right)= 0, \label{appB_eq0c} \\
&\frac{\partial \mathcal{L}}{\partial p(x_2|v)}=\frac{1}{2} \sum_{v \in \mathcal{V}}\log \left(1+\frac{v^2 P_{EHU}(x_2,v)}{\sigma_2^2 + x_2^2 \alpha_2}\right) p(v) - \lambda_1 \sum_{v \in \mathcal{V}}x_2^2 p(v) - \mu_1 \nonumber\\ &-\frac{\partial}{\partial p \left(x_2|v\right)} \left( \sum_{v \in \mathcal{V}}\sum_{g \in \mathcal{G}}\sum_{f \in \mathcal{F}} I(X_{1};Y_{3}|V=v, G=g, F=f)p(v)p(g)p(f)\right)  \nonumber\\
&-\lambda_2 \left((1-\eta(\bar {q_1}^2 +\alpha_1))\sum_{v \in \mathcal{V}}P_{EHU}(x_2,v) p(v)- \eta \sum_{v \in \mathcal{V}}v^2 x_2^2 p(v)\right)=0. \label{appB_eq2}
\end{align}

Now, when $P_{EHU}>0$, then $\mu_2=0$ in (\ref{appB_eq0c}). In consequence, we can use (\ref{appB_eq0c}) to find $P_{EHU}(x_2,v)$ as given by Theorem 2. If the solution is negative, then $P_{EHU}(x_2,v)=0$.

By using (\ref{appB_eq2}), we can prove that the optimal input probability distribution, $p(x_2|v)$, is discrete. The proof is based on (\cite{6193208}), where the authors derive a methodology which identifies the capacity-achieving distribution, based on standard decompositions
in Hilbert space with the Hermitian polynomials as a basis. Since
\begin{align}
&I(X_{1};Y_{3}|V=v, G=g, F=f)=H(Y_3 |E=e, G=g, F=f) \nonumber\\
&-H(Y_3|X_1=x_1,V=v, G=g, F=f)=I(X_{1};Y_{3}|V=v, G=g, F=f)\nonumber\\
&=H(Y_3 |V=v, G=g, F=f) -H(Z_3|V=v, G=g, F=f),
\end{align}
where $Z_3=G X_{2}+N_{3}$, first we note that
\begin{flalign} \label{app_subA_eq2}
&I'(X_{1};Y_{3}|V=v, G=g, F=f)\nonumber\\
&=H'(Y_3 |V=v, G=g, F=f) -H'(Z_3|V=v, G=g, F=f)\nonumber\\
&= \int_{-\infty}^{\infty} \frac{1}{\sqrt{2 \pi \sigma_{z_3}}} e^{-\frac{(z_3 - x_{2})^2}{2 \sigma_{z_3}^2}} \times \ln \left(p(z_3)\right) d z_3-\int_{-\infty}^{\infty} \frac{1}{\sqrt{2 \pi \sigma_{y_3}}} e^{-\frac{(y_3 - x_{2})^2}{2 \sigma_{y_3}^2}} \times \log \left(p(y_3)\right) d y_3.
\end{flalign}
where $'$ denotes the derivative with respect to $p(x_2|v)$. Now, we decompose the integrals in (\ref{app_subA_eq2}) by using Hermitian polynomials. To this end, we define
\begin{flalign} \label{app_subA_eq5}
\log(p(y_3))=\sum_{m=0}^{\infty} c^{(1)}_m H_m(y_3) \quad \text{and} \quad \log(p(z_3))=\sum_{m=0}^{\infty} c^{(2)}_m H_m(z_3),
\end{flalign}
where $c^{(1)}_m$ and $c^{(2)}_m$ are constants and $H_m(y_3)$ and $H_m(z_3)$ are the Hermitian polynomials, $\forall m$. When (\ref{app_subA_eq5}) is used in conjunction with the generating function of the Hermitian polynomials, given by
\begin{flalign} \label{app_subA_eq6}
e^{-\frac{t^2}{2}+tx}=\sum_{m=0}^{\infty} H_m(x)\frac{t^m}{m!},
\end{flalign}
for $H'(Y_3 |E=e, G=g, F=f)$ in (\ref{app_subA_eq2}) we obtain
\begin{flalign} \label{app_subA_eq7}
&H'(Y_3 |E=e, G=g, F=f)=-\int_{-\infty}^{\infty} \frac{1}{\sqrt{2 \pi}} e^{-\frac{(y_3 - x_{2})^2}{2 \sigma_{y_3}^2}} \sum_{m=0}^{\infty} c^{(1)}_m H_m(y_3) d y_3 \nonumber\\
&=-\int_{-\infty}^{\infty} \frac{1}{\sqrt{2 \pi}} e^{-\frac{y^2_3}{2}}e^{-\frac{x_2^2}{2}+x_2 y_3} \sum_{m=0}^{\infty} c^{(1)}_m H_m(y_3) d y_3 \nonumber\\
&=-\int_{-\infty}^{\infty} \frac{1}{\sqrt{2 \pi}} e^{-\frac{y^2_3}{2}}\sum_{n=0}^{\infty} H_n(x)\frac{t^n}{n!} \sum_{m=0}^{\infty} c^{(1)}_m H_m(y_3) d y_3 =-\sum_{m=0}^{\infty} c^{(1)}_m x_2^m.
\end{flalign}
In (\ref{app_subA_eq7}), we used the ortogonality of the Hermitian polynomials with respect to the weight function $e^{-\frac{y^2_3}{2}}$ and we set $\sigma_{y_3}^2=1$ for simplicity. By following an analogous procedure for $H'(Z_3 |E=e, G=g, F=f)$ in (\ref{app_subA_eq2}), we obtain
\begin{flalign} \label{app_subA_eq8}
H'(Z_3 |V=v, G=g, F=f)=-\sum_{m=0}^{\infty} c^{(2)}_m x_2^m.
\end{flalign}
In order to identify the constants $c^{(1)}_m$ and $c^{(2)}_m$ in (\ref{app_subA_eq5}), we consider 2 scenarios.

\underline{Case 1:} Let us assume $P_{EHU}(x_2,v) =0$. The condition given by (\ref{appB_eq2}) can be written as
\begin{flalign} \label{app_subA_eq9}
&\sum_{m=0}^{\infty} (c^{(1)}_m - c^{(2)}_m) x_2^m = \lambda_1 x_2^2 + \mu_1+\lambda_2 \left((1-\eta(\bar {q_1}^2 +\alpha_1))P_{EHU}- \eta e^2 x_2^2\right).
\end{flalign}
The comparison of the exponents of $x_2$ in (\ref{app_subA_eq9}) yields
\begin{flalign} \label{app_subA_eq10}
&c^{(1)}_0 = \mu_1, c^{(2)}_0 = 0; \quad c^{(1)}_1 = c^{(2)}_1 = 0; \quad c^{(1)}_2 = \lambda_1, c^{(2)}_2 = \lambda_1 \eta v^2; \quad c^{(1)}_m = c^{(2)}_m = 0, \forall m > 2.
\end{flalign}
Now, we can insert (\ref{app_subA_eq10}) into (\ref{app_subA_eq5}) and obtain
\begin{flalign} \label{app_subA_eq11}
p(y_3)=e^{\ln(2)(c^{(1)}_0 H_0 (y_3)+c^{(1)}_2 H_2 (y_3))}\stackrel{(a)}{=}e^{\ln(2)(c^{(1)}_0-c^{(1)}_2 )}e^{\ln(2)c^{(1)}_2 y_3^2},
\end{flalign}
where $(a)$ follows from the definition of Hermitian polynomials, i.e., $H_0 (y_3)=1$ and $H_2 (y_3)=y_3^2-1$. The expression given by (\ref{app_subA_eq11}) can only be a valid probability distribution iff $c^{(1)}_2 < 0$, in which case $p(y_3)$ would be distributed according to a normal distribution. Consequently, $x_2$ would also be a Gaussian RV. However, since $\lambda_1 \geq 0$, this would not be possible, thus $p(y_3)$ can not be a continuous probability distribution. A similar argument would follow for $p(z_3)$ in (\ref{app_subA_eq5}), and it would lead to an identical conclusion since $\lambda_1 \eta v^2$ can not be negative.

\underline{Case 2:} Let us assume $P_{EHU}(x_2,v) >0$. By using a Taylor series expansion we can rewrite the $\log(.)$ function in (\ref{appB_eq2}) as
\begin{align}\label{app_subA_eq12}
\frac{1}{2} \log \left(1+\frac{v^2 P_{EHU}(x_2,v)}{\sigma_2^2 + x_2^2 \alpha_2}\right) = \frac{1}{2}\sum_{n=0}^{\infty} (-1)^n a_n x_2^{2n},
\end{align}
therefore (\ref{appB_eq2}) can be written as
\begin{flalign} \label{app_subA_eq13}
&\sum_{m=0}^{\infty} (c^{(2)}_m - c^{(1)}_m) x_2^m = \frac{1}{2}\sum_{n=0}^{\infty} (-1)^n a_n x_2^{2n} - \lambda_1 x_2^2 - \mu_1-\lambda_2 \left((1-\eta(\bar {q_1}^2 +\alpha_1))P_{EHU}(x_2,v)- \eta v^2 x_2^2\right).
\end{flalign}
In (\ref{app_subA_eq12}) and (\ref{app_subA_eq13}), $a_n>0$ are known constants. By applying the same procedure as in Case 1, we obtain $c^{(2)}_m $ and $ c^{(1)}_m$ as
\begin{flalign} \label{app_subA_eq14}
&c^{(1)}_0 = \lambda_2 P_{EHU}(x_2,v) +\mu_1, c^{(2)}_0 = \frac{1}{2} a_n + \lambda_2 \eta (\bar {q_1}^2 +\alpha_1) P_{EHU}(x_2,v); \quad c^{(1)}_1 = c^{(2)}_1 = 0; \nonumber\\
&c^{(1)}_2 = \lambda_1, c^{(2)}_2 = \lambda_1 \eta v^2; \quad c^{(1)}_m =  0, c^{(2)}_m = \frac{1}{2} a_{m/2}, \quad \forall m > 2 \wedge \text{$m$ is even} \nonumber\\
&c^{(1)}_m = c^{(2)}_m = 0, \quad \forall m > 2 \wedge \text{$m$ is odd}.
\end{flalign}
Consequently,
\begin{flalign} \label{app_subA_eq15}
p(y_3)=e^{\ln(2)(c^{(1)}_0 H_0 (y_3)+c^{(1)}_2 H_2 (y_3))}\stackrel{(a)}{=}e^{\ln(2)(c^{(1)}_0-c^{(1)}_2 )}e^{\ln(2)c^{(1)}_2 y_3^2},
\end{flalign}
however, $\lambda_1 \geq 0$, so $c^{(1)}_2$ is positive, thus $p(y_3)$ can not be a valid continuous distribution. As for $p(z_3)$, we have
\begin{flalign} \label{app_subA_eq16}
p(z_3)=e^{\ln(2)\sum_{m=0}^{\infty} c^{(2)}_m H_m (z_3)}\stackrel{(a)}{=}e^{\ln(2)\sum_{n=0}^{\infty} q_n z_3^{2n}}=\prod_{n=0}^{\infty}e^{\ln(2)q_n z_3^{2n}},
\end{flalign}
where $(a)$ follows from the fact that $c^{(2)}_m>0$ only for even values of $m$ and $q_n$ are known non-zero constants, whose value is determined by the polynomials and $a_{n}$. Since $q_n > 0$ for some $n \to \infty$, $p(z_3)$ is unbounded, and as a result $p(x_2)$ can not be continues. Considering Case 1 and Case 2, we obtain that $p(x_2|v)$ has to be discrete on the entire domain of $x_2$. Now, we generate every discrete probability distribution satisfying C1 in (\ref{appB_eq0a}) and settle on the probability distribution which maximizes the secrecy rate.

In order to obtain $I(X_{1};Y_{3}|V=v, G=g, F=f)$, we use the definition of mutual information, and we can write
%\newpage
\begin{flalign} \label{proof_eq4}
&I(X_{1};Y_{3}|V=v, G=g, F=f) \nonumber\\
&=H(Y_{3}|V=v, G=g, F=f)- H(Y_{3}|X_{1}=x_1, V=v, G=g, F=f)  \nonumber\\
%&=H(Y_{3}|V=v, G=g, F=f)\nonumber\\
%&-H(F X_{1}+G X_{2}+N_{3}|X_{1}=x_1, V=v, G=g, F=f)  \nonumber\\
%&=H(Y_{3}|V=v, G=g, F=f) -H(\underbrace{G X_{2}+N_{3}}_{\text{$Z_3$}}|V=v, G=g, F=f)  \nonumber\\
&=\left( \int_{-\infty}^{\infty} \frac{1}{\sqrt{2 \pi \sigma_{y_3}}} \sum_{j=1}^{J} p (x_2 = x_{2j}) e^{-\frac{(y_3 - x_{2j})^2}{2 \sigma_{y_3}^2}} \times \ln \left(\frac{1}{\sqrt{2 \pi \sigma_{y_3}}} \sum_{j=1}^{J} p (x_2 = x_{2j}) e^{-\frac{(y_3 - x_{2j})^2}{2 \sigma_{y_3}^2}}\right) d y_3  \right.\nonumber\\
&-\int_{-\infty}^{\infty} \frac{1}{\sqrt{2 \pi \sigma_{3}}} \sum_{j=1}^{J} p (x_2 = x_{2j}) e^{-\frac{(z - x_{2j})^2}{2 \sigma_3^2}} \times \ln \left(\frac{1}{\sqrt{2 \pi \sigma_{3}}} \sum_{j=1}^{J} p (x_2 = x_{2j}) \left.e^{-\frac{(z - x_{2j})^2}{2 \sigma_3^2}}\right) d z_3 \right),
\end{flalign}
where the last equality is a consequence of the definition of entropy. Finally, by using (\ref{proof_eq4}) we obtain the upper bound as given in Theorem 2.

%------------------------ \bibliography--------------------------------------------%
\bibliography{litdab}

\begin{thebibliography}{10}
\providecommand{\url}[1]{#1}
\csname url@samestyle\endcsname
\providecommand{\newblock}{\relax}
\providecommand{\bibinfo}[2]{#2}
\providecommand{\BIBentrySTDinterwordspacing}{\spaceskip=0pt\relax}
\providecommand{\BIBentryALTinterwordstretchfactor}{4}
\providecommand{\BIBentryALTinterwordspacing}{\spaceskip=\fontdimen2\font plus
\BIBentryALTinterwordstretchfactor\fontdimen3\font minus
  \fontdimen4\font\relax}
\providecommand{\BIBforeignlanguage}[2]{{%
\expandafter\ifx\csname l@#1\endcsname\relax
\typeout{** WARNING: IEEEtran.bst: No hyphenation pattern has been}%
\typeout{** loaded for the language `#1'. Using the pattern for}%
\typeout{** the default language instead.}%
\else
\language=\csname l@#1\endcsname
\fi
#2}}
\providecommand{\BIBdecl}{\relax}
\BIBdecl

\bibitem{Sec_RFID}
W.~Saad, Z.~Han, and H.~V. Poor, ``On the physical layer security of
  backscatter rfid systems,'' in \emph{Proc. 2012 International Symposium on
  Wireless Communication Systems (ISWCS)}, Aug. 2012, pp. 1092--1096.

\bibitem{Sec_sensorNets}
J.~Choi, J.~Ha, and H.~Jeon, ``Physical layer security for wireless sensor
  networks,'' in \emph{Proc. 2013 IEEE 24th Annual International Symposium on
  Personal, Indoor, and Mobile Radio Communications (PIMRC)}, Sept. 2013, pp.
  1--6.

\bibitem{4626059}
P.~K. Gopala, L.~Lai, and H.~E. Gamal, ``On the secrecy capacity of fading
  channels,'' \emph{IEEE Trans. Inf. Theory}, vol.~54, no.~10, pp. 4687--4698,
  Oct. 2008.

\bibitem{SecCap_Shannon}
C.~E. Shannon, ``Communication theory of secrecy systems,'' \emph{The Bell
  System Technical Journal}, vol.~28, no.~4, pp. 656--715, Oct. 1949.

\bibitem{SecCap_Wyner}
A.~D. Wyner, ``The wire-tap channel,'' \emph{The Bell System Technical
  Journal}, vol.~54, pp. 1355–--1387, 1975.

\bibitem{SecCap_WPCNs}
H.~Ju and R.~Zhang, ``Throughput maximization in wireless powered communication
  networks,'' \emph{IEEE Trans. on Wireless Communications}, vol.~13, no.~1,
  pp. 418--428, Jan. 2014.

\bibitem{SecCap_EH}
D.~Gunduz, K.~Stamatiou, N.~Michelusi, and M.~Zorzi, ``Designing intelligent
  energy harvesting communication systems,'' \emph{IEEE Communications
  Magazine}, vol.~52, no.~1, pp. 210--216, Jan. 2014.

\bibitem{bi2016wireless}
S.~Bi, Y.~Zeng, and R.~Zhang, ``Wireless powered communication networks: An
  overview,'' \emph{IEEE Wireless Communications}, vol.~23, no.~2, pp. 10--18,
  April 2016.

\bibitem{liu2018exploiting}
Y.~Liu, J.~Xu, and R.~Zhang, ``Exploiting interference for secrecy wireless
  information and power transfer,'' \emph{IEEE Wireless Communications},
  vol.~25, no.~1, pp. 133--139, Feb. 2018.

\bibitem{SecCap_PowerConst}
C.~N. Nguyen, P.~C. Doan-Thi, D.~D. Tran, and D.~B. Ha, ``Secured energy
  harvesting networks with multiple power-constrained information sources,'' in
  \emph{Proc. 2017 International Conference on Recent Advances in Signal
  Processing, Telecommunications Computing (SigTelCom)}, Jan. 2017, pp.
  134--138.

\bibitem{SecCap_MIMO_Ulukus}
K.~Banawan and S.~Ulukus, ``Gaussian mimo wiretap channel under receiver side
  power constraints,'' in \emph{Proc. 2014 52nd Annual Allerton Conference on
  Communication, Control, and Computing (Allerton)}, Sept. 2014, pp. 183--190.

\bibitem{SecCap_MIMO_Ulukus_prim}
------, ``Mimo wiretap channel under receiver-side power constraints with
  applications to wireless power transfer and cognitive radio,'' \emph{IEEE
  Trans. on Communications}, vol.~64, no.~9, pp. 3872--3885, Sept. 2016.

\bibitem{SecCap_SIMO_SWIPT}
G.~Pan, C.~Tang, T.~Li, and Y.~Chen, ``Secrecy performance analysis for simo
  simultaneous wireless information and power transfer systems,'' \emph{IEEE
  Trans. on Communications}, vol.~63, no.~9, pp. 3423--3433, Sept. 2015.

\bibitem{liu2013secrecy}
L.~Liu, R.~Zhang, and K.-C. Chua, ``Secrecy wireless information and power
  transfer with miso beamforming,'' in \emph{Proc. Global communications
  conference (GLOBECOM), 2013 IEEE}, 2013, pp. 1831--1836.

\bibitem{SecCap_MISO_SWIPT}
G.~Pan, H.~Lei, Y.~Deng, L.~Fan, J.~Yang, Y.~Chen, and Z.~Ding, ``On secrecy
  performance of miso swipt systems with tas and imperfect csi,'' \emph{IEEE
  Trans. on Communications}, vol.~64, no.~9, pp. 3831--3843, Sept. 2016.

\bibitem{liu2017power}
M.~Liu and Y.~Liu, ``Power allocation for secure swipt systems with
  wireless-powered cooperative jamming,'' \emph{IEEE Communications Letters},
  vol.~21, no.~6, pp. 1353--1356, 2017.

\bibitem{zhang2016artificial}
M.~Zhang, Y.~Liu, and R.~Zhang, ``Artificial noise aided secrecy information
  and power transfer in ofdma systems,'' \emph{IEEE Transactions on Wireless
  Communications}, vol.~15, no.~4, pp. 3085--3096, 2016.

\bibitem{SecCap_Relay}
A.~Salem, K.~A. Hamdi, and K.~M. Rabie, ``Physical layer security with rf
  energy harvesting in af multi-antenna relaying networks,'' \emph{IEEE Trans.
  on Communications}, vol.~64, no.~7, pp. 3025--3038, July 2016.

\bibitem{SecCap_Jam}
T.~G. Nguyen, C.~So-In, D.-B. Ha \emph{et~al.}, ``Secrecy performance analysis
  of energy harvesting wireless sensor networks with a friendly jammer,''
  \emph{IEEE Access}, vol.~5, pp. 25\,196--25\,206, 2017.

\bibitem{SecCap_FD_Jam}
Y.~Bi and H.~Chen, ``Accumulate and jam: Towards secure communication via a
  wireless-powered full-duplex jammer,'' \emph{IEEE Journal of Selected Topics
  in Signal Processing}, vol.~10, no.~8, pp. 1538--1550, Dec. 2016.

\bibitem{SecCap_FD_Passive}
E.~Everett, A.~Sahai, and A.~Sabharwal, ``Passive self-interference suppression
  for full-duplex infrastructure nodes,'' \emph{IEEE Trans. on Wireless
  Communications}, vol.~13, no.~2, pp. 680--694, Feb. 2014.

\bibitem{SecCap_FD_Comb}
M.~Duarte, A.~Sabharwal, V.~Aggarwal, R.~Jana, K.~K. Ramakrishnan, C.~W. Rice,
  and N.~K. Shankaranarayanan, ``Design and characterization of a full-duplex
  multiantenna system for wifi networks,'' \emph{IEEE Trans. on Vehicular
  Technology}, vol.~63, no.~3, pp. 1160--1177, March 2014.

\bibitem{zhang2017secrecy}
C.~Zhang, D.~Wang, J.~Ye, H.~Lei, J.~Zhang, G.~Pan, and Q.~Feng, ``Secrecy
  outage analysis on underlay cognitive radio system with full-duplex secondary
  user,'' \emph{IEEE Access}, vol.~5, pp. 25\,696--25\,705, 2017.

\bibitem{tang2017physical}
W.~Tang, S.~Feng, Y.~Ding, and Y.~Liu, ``Physical layer security in
  heterogeneous networks with jammer selection and full-duplex users,''
  \emph{IEEE Transactions on Wireless Communications}, vol.~16, no.~12, pp.
  7982--7995, 2017.

\bibitem{SecCap_FD_TripleSE}
S.~Haddad, A.~{\"O}zg{\"u}r, and E.~Telatar, ``Can full-duplex more than double
  the capacity of wireless networks?'' in \emph{Proc. Information Theory
  (ISIT), 2017 IEEE International Symposium on}, 2017, pp. 963--967.

\bibitem{nikoloska2018capacity}
I.~Nikoloska, N.~Zlatanov, and Z.~Hadzi-Velkov, ``Capacity of a full-duplex
  wirelessly powered communication system with self-interference and processing
  cost,'' \emph{IEEE Transactions on Wireless Communications}, vol.~17, no.~11,
  pp. 7648--7660, 2018.

\bibitem{FD_SI_recycling}
Y.~Zeng and R.~Zhang, ``Full-duplex wireless-powered relay with self-energy
  recycling,'' \emph{IEEE Wireless Communications Letters}, vol.~4, no.~2, pp.
  201--204, April 2015.

\bibitem{SecCap_FDradios}
D.~Bharadia, E.~McMilin, and S.~Katti, ``Full duplex radios,'' \emph{SIGCOMM
  Comput. Commun. Rev.}, vol.~43, no.~4, Aug.

\bibitem{SecCap_zlatanov}
N.~Zlatanov, E.~Sippel, V.~Jamali, and R.~Schober, ``Capacity of the gaussian
  two-hop full-duplex relay channel with residual self-interference,''
  \emph{IEEE Trans. on Communications}, vol.~65, no.~3, pp. 1005--1021, March
  2017.

\bibitem{cover2012elements}
T.~M. Cover and J.~A. Thomas, \emph{{Elements of Information Theory}}.\hskip
  1em plus 0.5em minus 0.4em\relax John Wiley \& Sons, 2012.

\bibitem{xu2014throughput}
J.~Xu and R.~Zhang, ``Throughput optimal policies for energy harvesting
  wireless transmitters with non-ideal circuit power,'' \emph{IEEE Journal on
  Selected Areas in Communications}, vol.~32, no.~2, pp. 322--332, Feb. 2014.

\bibitem{eh_zlatanov}
N.~Zlatanov, R.~Schober, and Z.~Hadzi-Velkov, ``Asymptotically optimal power
  allocation for energy harvesting communication networks,'' \emph{IEEE Trans.
  on Vehicular Technology}, vol.~66, no.~8, pp. 7286--7301, Aug 2017.

\bibitem{e10030200}
J.~V. Michalowicz, J.~M. Nichols, and F.~Bucholtz, ``Calculation of
  differential entropy for a mixed gaussian distribution,'' \emph{Entropy},
  vol.~10, no.~3, pp. 200--206, 2008.

\bibitem{cover}
T.~Cover and A.~{El Gamal}, ``{Capacity theorems for the relay channel},''
  \emph{IEEE Trans. Inf. Theory}, vol.~25, pp. 572--584, Sep. 1979.

\bibitem{Boyd_CO}
S.~Boyd and L.~Vandenberghe, \emph{Convex Optimization}.\hskip 1em plus 0.5em
  minus 0.4em\relax Cambridge University Press, 2004.

\bibitem{6193208}
J.~Fahs and I.~Abou-Faycal, ``{Using hermite bases in studying
  capacity-achieving distributions over AWGN channels},'' \emph{IEEE Trans.
  Inf. Theory}, vol.~58, pp. 5302--5322, Aug. 2012.

\end{thebibliography}
\bibliographystyle{IEEEtran}
\end{document}